\documentclass[aps,twocolumn,pra]{revtex4}

\usepackage{amsmath,amssymb,bm}
\usepackage{graphicx}
\usepackage{epstopdf}
\usepackage{latexsym}
\usepackage{subfigure}
\usepackage[usenames,dvipsnames]{color}
\usepackage{hyperref}
\usepackage{natbib}
\usepackage{moreverb}

\begin{document}
\newcommand{\cs}[1]{{\color{blue}$\clubsuit$#1}}
\newcommand{\sinc}{\mathrm{sinc}}
\newcommand{\red}[1]{{ \bf \color{red}#1}}

\title{Effective many-body parameters for atoms in non-separable Gaussian optical potentials}
\author{Michael L. Wall $^{1}$\email{mwall.physics@gmail.com}, Kaden R.~A.~Hazzard$^{2}$, A.~M. Rey$^{1}$ }
\affiliation{$^{1}$JILA, NIST, Department of Physics, University of Colorado, 440 UCB, Boulder, CO 80309, USA}
\affiliation{$^{2}$ Department of Physics, Rice University, Houston, Texas 77005, USA}
\date{\today}

\begin{abstract}
We analyze the properties of particles trapped in three-dimensional potentials formed from superimposed Gaussian beams, fully taking into account effects of potential anharmonicity and non-separability.  Although these effects are negligible in more conventional optical lattice experiments, they are essential for emerging ultracold atom developments.  We focus in particular on two potentials utilized in current ultracold atom experiments: arrays of tightly focused optical tweezers and a one-dimensional optical lattice with transverse Gaussian confinement and highly excited transverse modes.  Our main numerical tools are discrete variable representations (DVRs), which combine many favorable features of spectral and grid-based methods, such as the computational advantage of exponential convergence and the convenience of an analytical representation of Hamiltonian matrix elements.  Optimizations, such as symmetry adaptations and variational methods built on top of DVR methods, are presented and their convergence properties discussed.  We also present a quantitative analysis of the degree of non-separability of eigenstates, borrowing ideas from the theory of matrix product states (MPSs), leading to both conceptual and computational gains.  Beyond developing numerical methodologies, we present results for construction of optimally localized Wannier functions and tunneling and interaction matrix elements in optical lattices and tweezers relevant for constructing effective models for many-body physics.

\end{abstract}
\maketitle

\section{Introduction}
\label{sec:introduction}

Ultracold neutral atoms trapped in optical potentials have been solidly established as a highly controllable platform for precision measurement and quantum metrology~\cite{doi:10.1142/S0218271807011826} as well as quantum simulation of many-body physics~\cite{bloch2008many,bloch2012quantum}.  The most prevalent method for quantum simulation of three-dimensional (3D) lattice systems is to optically trap neutral atoms in a cubic lattice formed by three orthogonal pairs of counterpropagating laser beams~\cite{bloch2005ultracold}.  An emerging alternative for manipulating and trapping ultracold atoms is optical tweezers~\cite{Schlosser2001}, in which a tightly focused Gaussian beam is used to trap single atoms.  In contrast to optical lattice experiments, which typically load gases which have been evaporatively cooled, atoms in optical tweezers can be individually manipulated and cooled to their ground state using laser cooling alone~\cite{Kaufman2012}.  The ability to dynamically vary the position of such tweezers at separations where tunneling is appreciable, as has been demonstrated in recent experiments~\cite{Kaufman_Lester_14,Jochim}, leads to a ``bottom-up" approach to building low-entropy quantum systems, in contrast to the ``top-down" approach of most optical lattice experiments in which large ensembles of atoms are cooled in an external trap before being loaded into the lattice.

The many-body physics of ultracold atoms in optical potentials is usually described within the framework of Hubbard models, which are truncated expansions of the full Hamiltonian in a basis of spatially localized orbitals known as Wannier functions~\cite{A&M}.  The parameters appearing in such models, for example tunneling integrals and interaction matrix elements, are the point of connection between few-body physics in the confining potential and many-body physics: these parameters controlling the many-body physics can be determined from few-body calculations or experiments.  Hence, determining them quantitatively is key to designing and validating robust quantum simulators and other quantum technologies.  The construction of Hubbard models for cubic optical lattices is greatly facilitated by the fact that such lattices are separable in Cartesian coordinates, $V(\mathbf{r})=V(x)+V(y)+V(z)$, and hence their eigenfunctions are products of eigenfunctions of one-dimensional (1D) problems, which are much less computationally demanding than 3D problems.  In addition, the theory of periodic potentials in 1D is especially simple and leads to computationally efficient procedures for determining the Wannier functions with maximum spatial localization~\cite{Kohn_59}.  On the other hand, technologies which depend upon the curvature of light beams for trapping, such as optical tweezers, are inherently non-separable, and so the well-developed machinery employed for cubic lattices is not applicable.  In addition to being more computationally challenging, non-separable potentials have significant qualitative differences from separable potentials.  For example, the tunneling rate of a particle through a non-separable lattice depends on its transverse motional state, and can change this rate by an order of magnitude or more.  In contrast, for a separable potential tunneling rates are independent of the transverse motional state.

In this work, we study the eigenstates of non-separable 3D optical potentials constructed by superimposing Gaussian laser beams, taking as our two main examples arrays of optical tweezers~\cite{Kaufman_Lester_14,Jochim} and a 1D optical lattice with transverse Gaussian confinement, such as is utilized for optical lattice clocks~\cite{Martin09082013}.  Our main numerical tool is discrete variable representations (DVRs), coupled with variational methods.  As will be discussed in Sec.~\ref{sec:DVR}, DVR methods combine many nice features of grid-based methods, such as analytic representation of the kinetic energy operator and a diagonal representation of the potential energy operator, with features of spectral methods, such as exponential convergence.  We focus in particular on the connections between the trap parameters, e.g., the trap depth, and the parameters appearing in effective many-body models, such as tunneling and interaction matrix elements.  We also quantitatively study the degree of separability of the eigenfunctions of such potentials using tools from quantum information science.  In particular, borrowing from the theory of matrix product states (MPSs)~\cite{Schollwoeck_11}, which accurately capture many-body states with restricted entanglement, we develop an analogous state ansatz for non-separable single-particle states.  As we will show, this ansatz is useful for visualization, storage, and computational efficiency.

This work is organized as follows: Sec.~\ref{eq:OptPot} defines the potentials we study in this work, their symmetries, and the Hubbard parameters we will study; Sec.~\ref{sec:Numerical} reviews the two DVR basis sets we use and presents numerical optimizations for DVR-based algorithms; Sec.~\ref{sec:NearestSeparable} presents an analysis of non-separable single-particle states from the viewpoint of quantum information theory, in particular discussing a canonical form for non-separable states motivated by matrix product states; Sec.~\ref{sec:Restuls} gives results for Hubbard parameters and quantifies non-separability of states in a double-well optical tweezer array and a non-separable optical lattice; Finally, Sec.~\ref{sec:Concl} concludes and gives an outlook.  Some technical details on computing interaction matrix elements in a basis of radial functions and a variational algorithm for finding the nearest separable state given a state in the MPS canonical form are given as appendices.

\section{Optical potentials for neutral atoms}
\label{eq:OptPot}

\begin{figure}[!tbp]
\centerline{\includegraphics[width=0.7 \columnwidth]{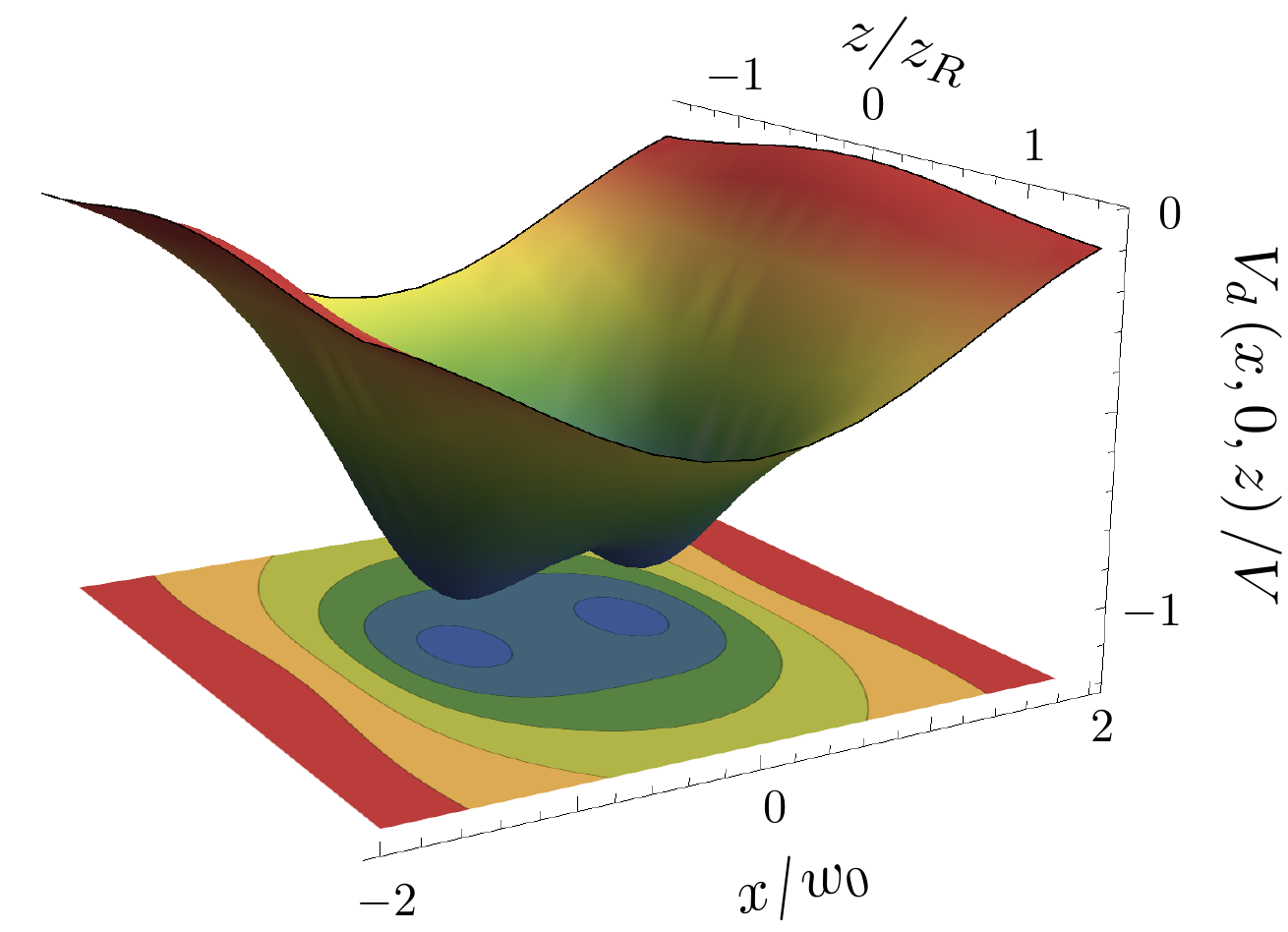}}
\caption{\label{fig:TweezFig} 
\emph{{Double-well optical tweezer.}} The double-well optical tweezer potential, Eq.~\eqref{eq:dTweez}, at $y=0$, with no bias ($\Delta=0$) and spacing $a/w_0=1.23$.  Note the different scaling of the $x$ and $z$ coordinates.}
\end{figure}

In this work, we consider optical potentials which are generated by superimposed Gaussian laser beams. The electric field amplitude of the fundamental TEM$_{00}$ mode of a Gaussian laser beam propagating along $z$ may be written as~\cite{Svelto}
\begin{align}
\label{eq:EGaussian} |E\left(r,z\right)|&=\frac{E_0}{\sqrt{1+(\frac{z}{z_R})^2}}\exp\left(-\frac{r^2}{w_0^2(1+(\frac{z}{z_R})^2)}\right)\, ,
\end{align}
where $E_0$ is the field amplitude, the beam waist $w_0$ is where the field amplitude drops to $1/e$ of its on-axis value, and the Rayleigh range $z_R=\pi w_0^2/\lambda$, with $\lambda$ the wavelength of the laser light.  The resulting optical potential is proportional to the intensity of the field and the atomic polarizability.  For a single field of the form Eq.~\eqref{eq:EGaussian} with the laser-frequency far detuned from an optical transition, the ac-Stark shift gives rise to the potential
\begin{align}
\label{eq:sTweez} V_s\left(\mathbf{r}\right)&=-\frac{V}{1+\frac{z^2}{z_R^2}}\exp\left(\frac{-2r^2}{w_0^2\left(1+\frac{z^2}{z_R^2}\right)}\right)\, ,
\end{align}
with maximum depth at the origin $V$.  We assume the laser is red-detuned, and therefore that $V$ is positive.  We will refer to a single potential of the form Eq.~\eqref{eq:sTweez} as an optical tweezer when the beam waist is comparable to the wavelength $w_0\simeq \lambda$.  Such a potential can be formed by focusing a Gaussian beam through a high numerical aperture lens.  Expanding Eq.~\eqref{eq:sTweez} to lowest order in $r$ and $z$ results in a harmonic approximation
\begin{align}
V_s\left(\mathbf{r}\right)&\approx-V+\frac{2V}{w_0^2}r^2+\frac{V}{z_R^2}z^2\, ,
\end{align}
corresponding to radial and axial frequencies $\hbar\omega_r=\sqrt{8 E_{w_0} V}$ and $\hbar\omega_z=\sqrt{4E_{z_R}V}$, respectively, where $E_{w_0}=\hbar^2/2mw_0^2$ and $E_{z_R}=\hbar^2/2mz_R^2$ are the characteristic energies associated with the waist and Rayleigh range, respectively.   

The second case we consider is that of two focused traps in which one has an intensity greater by $\Delta$ more than the other, resulting in a double-well optical tweezer potential of the form 
\begin{widetext}
\begin{align}
\label{eq:dTweez} V_d\left(\mathbf{r}\right)&=-\frac{1}{1+\frac{z^2}{z_R^2}}\exp\left(\frac{-2y^2}{w_0^2\left(1+\frac{z^2}{z_R^2}\right)}\right)\left[V\exp\left(-\frac{2\left(x+a/2\right)^2}{w_0^2\left(1+\frac{z^2}{z_R^2}\right)}\right)+\left(V+\Delta\right) \exp\left(-\frac{2\left(x-a/2\right)^2}{w_0^2\left(1+\frac{z^2}{z_R^2}\right)}\right)\right]\, ,
\end{align}
\end{widetext}
as shown in Fig.~\ref{fig:TweezFig}, when the lasers are set up to avoid coherent interference effects, for example by frequency detuning the two beams with a splitting above any trap time scales.  The dynamics of two bosonic~\cite{Kaufman_Lester_14} or fermionic~\cite{Jochim} atoms in such a double-well tweezer configuration have been investigated in recent experiments.

The third case that we consider is a 1D optical lattice with transverse Gaussian character, which can be created by interfering two counter-propagating beams of the form Eq.~\eqref{eq:EGaussian}.  Confinement along the $z$ direction is provided by the standing wave interference pattern, and so $w_0\simeq \lambda$ is not required for 3D confinement.  Hence, we consider the regime $w_0\gg \lambda$ in which most experiments operate.  In this regime, we can neglect any effects of the Rayleigh range, giving rise to the potential
\begin{align}
\label{eq:Vlatt}V\left(r,z\right)&=-V_{\mathrm{latt}}\exp\left(-\frac{2r^2}{w_0^2}\right)\cos^2\left(k z\right)\, ,
\end{align}
where $k=\pi/a$, with $a=\lambda/2$ the lattice spacing.  In addition, to study the effects of non-separable transverse confinement on the axial motion of particles through a 1D optical lattice, we will consider the generalized 1D optical lattice
\begin{align}
\label{eq:VlattX}V_{\mathrm{latt}}\left(r,z\right)&=-\exp\left(-\frac{2r^2}{w_0^2}\right)\left[V_{\mathrm{const}}+V\cos^2\left(k z\right)\right]\, ,
\end{align}
which is comprised of a standing wave optical lattice Eq.~\eqref{eq:Vlatt} together with the potential resulting from a non-reflected beam Eq.~\eqref{eq:sTweez} of the same waist.  In this configuration, the transverse confinement frequency is set by $\sqrt{8 E_{w_0} \left(V_{\mathrm{const}}+V\right)}$, while confinement along $z$ measured close to a single lattice minimum is set by $V$ alone.  
\begin{figure}[tbp]
\centerline{\includegraphics[width=0.7 \columnwidth]{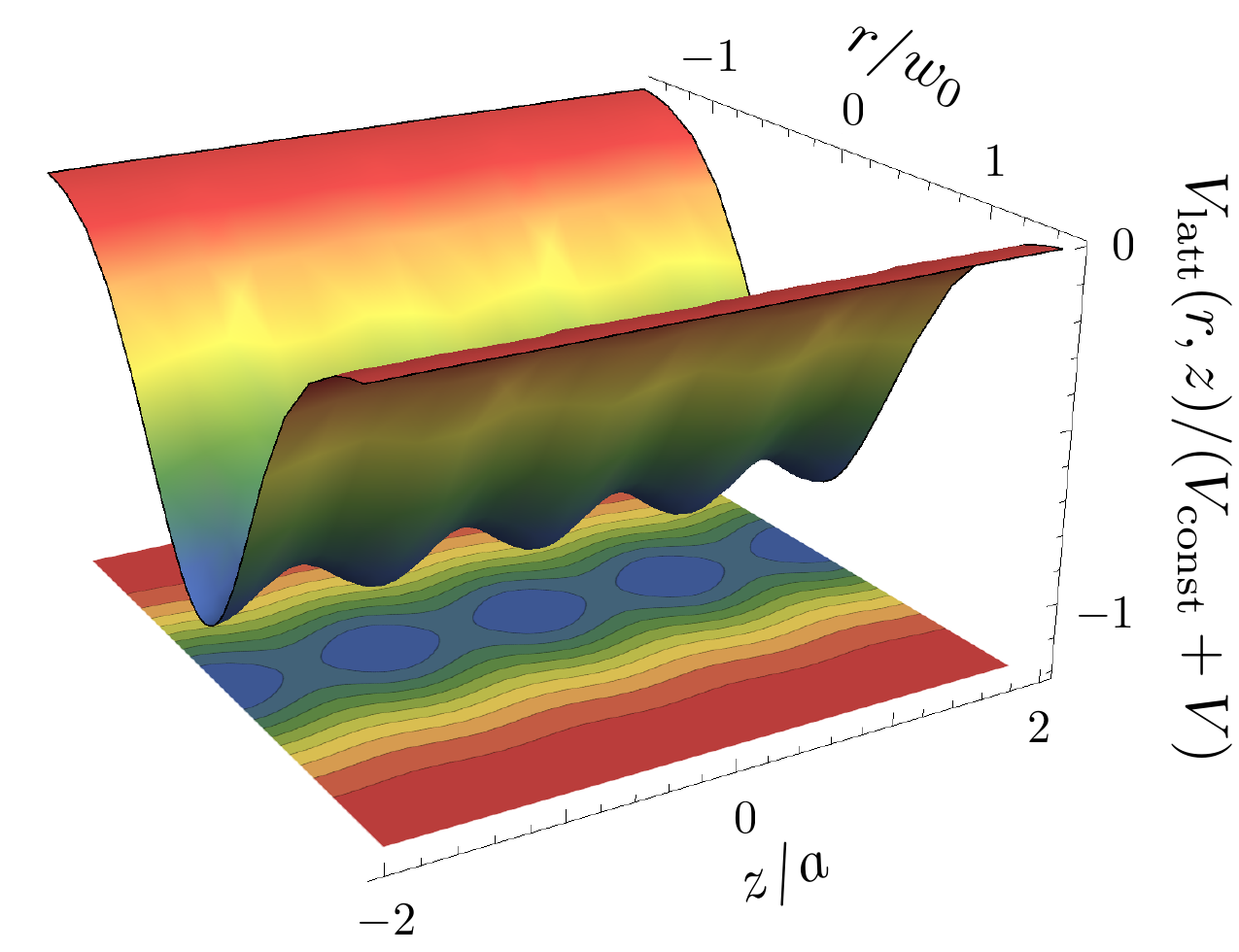}}
\caption{\label{fig:LattFig} 
\emph{{Non-separable optical lattice.}} The non-separable optical lattice, Eq.~\eqref{eq:VlattX}, for $V/V_{\mathrm{const}}=3/22$.  A non-zero value of $V_{\mathrm{const}}$ increases the effective transverse ($r$) confinement while leaving the axial ($z$) motion unchanged, to lowest order.  Note the different scalings of the $r$ and $z$ coordinates, as well as the difference in scaling of $z$ coordinates with respect to Fig.~\ref{fig:TweezFig} due to the additional length scale $a$.}
\end{figure}

Atoms loaded into an optical tweezer potential can be efficiently cooled to their 3D ground state via laser cooling~\cite{Kaufman2012}, and so our analysis of tweezer potentials will be focused on the properties of the lowest few eigenstates.  In contrast, some applications of optical lattices, such as optical atomic clocks~\cite{Martin09082013,Hinkley13092013}, do not require being in the lowest transverse motional states.   Hence, in discussing the 1D optical lattice potential we will devote special attention to develop methods which can efficiently deal with a large number of states to facilitate thermal averages.

\subsection{Symmetries of the optical potentials and characterization of eigenstates}
For a periodic potential such as the 1D optical lattice potentials Eq.~\eqref{eq:Vlatt}-\eqref{eq:VlattX}, the solutions may be characterized by a quasimomentum $q$ in the first Brillouin Zone (BZ), $q\in\left[-\pi/a,\pi/a\right)$, even in the case that the potential is non-separable.  The resulting eigenfunctions can be written in Bloch form $\psi_{nq}\left(\mathbf{r}\right)=e^{i qz}u_{nq}\left(\mathbf{r}\right)/\sqrt{L}$, where $u_{nq}\left(x,y,z+a\right)=u_{nq}\left(x,y,z\right)$ is a unit periodic function in $z$, with $a$ the lattice spacing and $L$ the number of unit cells.  Eigenfunctions of different $q$ are orthogonal by construction, and so can be obtained in separate calculations.  The double-well potential Eq.~\eqref{eq:dTweez} has a mirror symmetry along the $y$ and $z$ directions.  As a consequence, the solutions can be characterized in terms of their parities $P_y$ and $P_z$, where $P_{\nu}=1$ $(-1)$ corresponds to a function which is even (odd) under inversion of coordinate $\nu$.  In addition, for $\Delta=0$ there is also mirror symmetry along $x$.  As with the quasimomentum, we can obtain the eigenstates in each symmetry sector separately, leading to significant computational gains.

For a non-separable potential, we cannot unambiguously assign quantum numbers counting the number of quanta of excitation along each direction.  However, in most cases it is possible to assign labels which specify that the separable state \emph{nearest} to the given state may be labeled by a vector of excitation quanta $\mathbf{n}=\left(n_r,n_z\right)$ for problems of cylindrical symmetry and $\mathbf{n}=(n_x,n_y,n_z)$ for problems without cylindrical symmetry.  This vector $\mathbf{n}$ can be determined by counting the number of nodes in the components of the nearest separable wavefunction.  A precise definition of the nearest separable state and a procedure for obtaining it are presented later in Sec.~\ref{sec:NearestSeparable}.

In summary, the eigenstates of the optical lattice potentials Eqs.~\eqref{eq:Vlatt} and \eqref{eq:VlattX} can be written as $\psi_{m n_rn_zq}\left(\mathbf{r}\right)$, where $q$ is the quasimomentum, $m$ is the azimuthal quantum number arising from cylindrical symmetry, and $n_r$ and $n_z$ are labels stating that the state has character dominated by $n_r$ radial excitation quanta and $n_z$ band excitation quanta.  Similarly, eigenfunctions of the double-well tweezer potential Eq.~\eqref{eq:dTweez} may be written as $\psi_{\mathbf{n},\mathbf{p}}$, where $\mathbf{p}$ is a vector of parities and $\mathbf{n}$ a vector of excitation quanta labels.

\subsection{Hubbard parameters}

The transition from few- to many-particle physics in the presence of a trapping potential is frequently done by means of a Hubbard-type model, which projects the full many-body model onto a basis of low-energy lattice states.  Such a projection often removes irrelevant degrees of freedom without significantly modifying the physical behavior, resulting in models which are easier to analyze.  The many-body Hamiltonian in second quantization is
\begin{align}
\hat{H}=&\int d\mathbf{r} \hat{\psi}^{\dagger}\left(\mathbf{r}\right)\left[-\frac{\hbar^2}{2M} \nabla^2+V\left(\mathbf{r}\right)\right] \hat{\psi}\left(\mathbf{r}\right)\\
\nonumber &+\frac{1}{2}\int d\mathbf{r}d\mathbf{r}'\hat{\psi}^{\dagger}\left(\mathbf{r}\right)\hat{\psi}^{\dagger}\left(\mathbf{r}'\right)\hat{H}_{\mathrm{int}}\left(\mathbf{r}-\mathbf{r}'\right)\hat{\psi}\left(\mathbf{r}'\right)\hat{\psi}\left(\mathbf{r}\right)\, ,
\end{align}
where the first line represents single-particle physics in the trapping potential $V\left(\mathbf{r}\right)$ and the second line is two-body interactions with interaction Hamiltonian $\hat{H}_{\mathrm{int}}\left(\mathbf{r}\right)$.  This Hamiltonian can be expressed in terms of a particular single-particle basis $\{\psi_{\nu}\left(\mathbf{r}\right)\}$ by expanding the field operators in terms of this basis, $\hat{\psi}\left(\mathbf{r}\right)=\sum_{\nu} \psi_{\nu}\left(\mathbf{r}\right)\hat{a}_{\nu}$, where $\hat{a}_{\nu}$ is an operator which destroys a particle in state $\nu$.  A Hubbard model results when this expansion is not complete, but runs only over low-energy states in the single-particle basis.  These basis functions are usually taken to be localized Wannier functions $\{w_{i\mu}\left(\mathbf{r}\right)\}$, where $i$ denotes the lattice site (or potential minimum, in the case of a double-well optical tweezer) where the function is centered, and $\mu$ are any other single-particle quantum numbers or labels.  The precise construction of Wannier functions we use will be discussed in Sec.~\ref{sec:Restuls}.  A primary reason to choose the Wannier orbitals as the basis is that they decay exponentially in space and the couplings can therefore be truncated, for example including only nearest-neighbor tunneling in the tight-binding approximation.  The single-particle contributions to the Hubbard model with these approximations can be written as
\begin{align}
\label{eq:H1}\hat{H}_1&=\sum_{\mu} \sum_{i} E_{\mu}\hat{n}_{i\mu}-\sum_{\mu}J_{\mu}\sum_{\langle i,j\rangle}[\hat{a}^{\dagger}_{i\mu}\hat{a}_{j\mu}+\mathrm{h.c.}]\, ,
\end{align}
where $\mu$ runs over the restricted subset of single-particle states, $i$ runs over all lattice sites, $\langle i,j\rangle$ denotes a sum over neighboring lattice sites $i$ and $j$, $\hat{a}^{\dagger}_{ i\mu}$ creates a particle in state $\mu$ at lattice site $i$, and
\begin{align}
\label{eq:onsiteE}E_{\mu}&=\int d\mathbf{r} w_{i\mu}^{\star}\left(\mathbf{r}\right)\left[-\frac{\hbar^2}{2M} \nabla^2+V\left(\mathbf{r}\right)\right]w_{i\mu}\left(\mathbf{r}\right)\, ,\\
\label{eq:tunn}J_{\mu}&=-\int d\mathbf{r} w_{i\mu}^{\star}\left(\mathbf{r}\right)\left[-\frac{\hbar^2}{2M} \nabla^2+V\left(\mathbf{r}\right)\right]w_{j\mu}\left(\mathbf{r}\right)\, ,
\end{align}
with $i$ and $j$ nearest neighbors, are the on-site energies and tunneling matrix elements, respectively.

At ultracold temperatures, only the lowest partial waves contribute to scattering.  For neutral atoms, interactions can be well-modeled by zero-range pseudopotentials:  an $s$-wave pseudopotential for bosons
\begin{align}
\hat{H}_{\mathrm{s-wave}}\left(\mathbf{r}\right)&=\frac{4\pi \hbar^2 a_s}{M} \delta\left(\mathbf{r}\right)\partial_r r
\end{align}
and a $p$-wave pseudopotential for identical fermions~\cite{Idziaszek_Calarco_06}
\begin{align}
\hat{H}_{\mathrm{p-wave}}\left(\mathbf{r}\right)&=\frac{\pi \hbar^2 b_p^3}{M}\overleftarrow{\nabla} \delta\left(\mathbf{r}\right)\overrightarrow{\nabla} r \partial_{rrr} r^2\, .
\end{align}
Here, $a_s$ is the $s$-wave scattering length, $b_p^3$ the $p$-wave scattering volume, and the arrows on the $\nabla$ operators denote the direction of operation.  Keeping only interactions between particles on the same lattice site, the interaction contribution to the Hubbard model can be written
\begin{align}
\nonumber \label{eq:H2}\hat{H}_2=&\frac{1}{2}\sum_{\mu_1'\mu_2'; \mu_2\mu_1}\left(\frac{4\pi \hbar^2 a_s}{M}U_{\mu_1'\mu_2';\mu_2\mu_1}+\frac{6\pi \hbar^2 b_p^3}{M}V_{\mu_1'\mu_2';\mu_2\mu_1}\right)\\
&\times \sum_i \hat{a}^{\dagger}_{i\mu_1'}\hat{a}^{\dagger}_{i\mu_2'}\hat{a}_{i\mu_2}\hat{a}_{i\mu_1}\, ,
\end{align}
where the $s$- and $p$-wave Wannier integrals are
\begin{align}
\label{eq:swavepp} &U_{\mu_1'\mu_2';\mu_2\mu_1}=\int d\mathbf{r}w_{i\mu_1'}^{\star}\left(\mathbf{r}\right)w_{i\mu_2'}^{\star}\left(\mathbf{r}\right)w_{i\mu_2}\left(\mathbf{r}\right)w_{i\mu_1}\left(\mathbf{r}\right)\, ,\\
\nonumber  &V_{\mu_1'\mu_2';\mu_2\mu_1}=\int d\mathbf{r}\\
\nonumber & \times  \left[w_{i\mu_1'}^{\star}\left(\mathbf{r}\right)\left(\nabla w_{i\mu_2'}^{\star}\left(\mathbf{r}\right)\right)-\left(\nabla w_{i\mu_1'}^{\star}\left(\mathbf{r}\right)\right)w_{i\mu_2'}^{\star}\left(\mathbf{r}\right)\right]\\
\label{eq:pwavepp}&\cdot \left[\left(\nabla w_{i\mu_1}^{\star}\left(\mathbf{r}\right)\right)w_{i\mu_2}^{\star}\left(\mathbf{r}\right)-w_{i\mu_1}^{\star}\left(\mathbf{r}\right)\left(\nabla w_{i\mu_2}^{\star}\left(\mathbf{r}\right)\right)\right]\, .
\end{align}
Thus, determining the Wannier functions defines an effective Hubbard model $H=H_1+H_2$ with $H_1$ and $H_2$ given by Eqs.~\eqref{eq:H1} and \eqref{eq:H2}.

\section{Numerical Methods}
\label{sec:Numerical}
In this section we discuss the numerical methods we use to obtain the Wannier functions and, from these, the effective Hubbard parameters for particles trapped in non-separable potentials.  In particular, we advocate the use of discrete variable representations (DVRs) as a flexible, simple, and efficient means of solving the single-particle problem in anharmonic optical potentials.  This choice was motivated by the desire to have the very rapid convergence of a spectral method while still maintaining the flexibility and simplicity of grid-based methods.  Rapid convergence is desired both to reduce the computational demand of solving fully three-dimensional problems and also so that possibly a large number of states can be accurately converged with modest resources.

\subsection{Discrete variable representations}
\label{sec:DVR}

In this section, we briefly review the theory of the two DVRs we employ in this work: the sinc DVR and the Bessel DVR.  While the idea of DVRs is quite old, it was not until their apparent rediscovery in the 1980s that they found broad applicability in chemical and molecular physics (see e.g.~\cite{Light_review} for a review of this history).  In spite of their widespread use in chemical physics, DVRs have received less attention in the ultracold gases community (however, see~\cite{Nygaard_PRA,Nygaard_PRL}).  Perhaps the greatest advantage of DVR methods compared to pure spectral methods is their simplicity.  As will be shown below, applying DVR methods requires only the diagonalization of a matrix whose elements are all analytically known, as contrasted with spectral methods in which the Hamiltonian matrix elements consist of integrals which either have to be performed analytically for each problem instance or approximated numerically.  The advantage of DVR methods over other grid-based methods, such as finite-order finite differencing schemes, is in efficiency.  Standard finite-differencing schemes have an error which scales as a power of the number of grid points, while DVR methods converge exponentially in this parameter.

We define a DVR as follows~\cite{Littlejohn_Cargo_General}.  Given a domain $\mathcal{D}\subseteq \mathbb{R}$ and a Hilbert space $\mathcal{H}$ of functions on $\mathcal{D}$, we would like to find a subspace $\mathcal{S}_K$ of $\mathcal{H}$ with refinement parameter $K$ which captures the part of $\mathcal{H}$ spanned by low-energy eigenstates of our potential.  Let $\mathcal{P}_K$ be a Hermitian, idempotent projection operator into the subspace $\mathcal{S}_K$, and pick some set of $N$ grid points $\left\{x_n\right\}$, where $N=\mathrm{dim}\,\mathcal{S}_K$\footnote{In both the DVRs we define below, the spaces $\mathcal{S}_K$ are infinite dimensional, and so are not captured fully by a basis with a finite dimension $N$.  In these cases, we understand $\mathcal{S}_K$ to be the finite-dimensional subspace which is spanned by the $N$ DVR basis functions.}.  We then call the set of $\mathcal{P}_K$ and $\left\{x_n\right\}$ a discrete variable representation if the projected delta functions $|x_n\rangle=\mathcal{P}_K\delta\left(x-x_n\right)$ are orthogonal.  A normalized set of these projected delta functions, which we will denote in Dirac notation as $|\Delta_n\rangle =\mathcal{P}_K|x_n\rangle/\sqrt{N_n}$, with $N_n$ a normalization factor, becomes the set of basis functions in which we expand our wave functions of interest.  Methods which use an expansion in terms of basis functions defined on a grid in real space are also known as collocation-spectral methods or pseudospectral methods~\cite{PS}.  DVRs combine several nice properties of grid-based and spectral methods, such as exponential convergence in $N$ of eigenenergies and eigenvectors for problems with potentials $V\left(x\right)\in \mathcal{S}_K$, a diagonal representation for the potential energy $\langle \Delta_n|V\left(x\right)|\Delta_{n'}\rangle\approx V\left(x_n\right)\delta_{n,n'}$, and an analytic representation of the kinetic energy operator.

\subsubsection{Sinc DVR}

\begin{figure}[tbp]
\centerline{\includegraphics[width=0.8 \columnwidth]{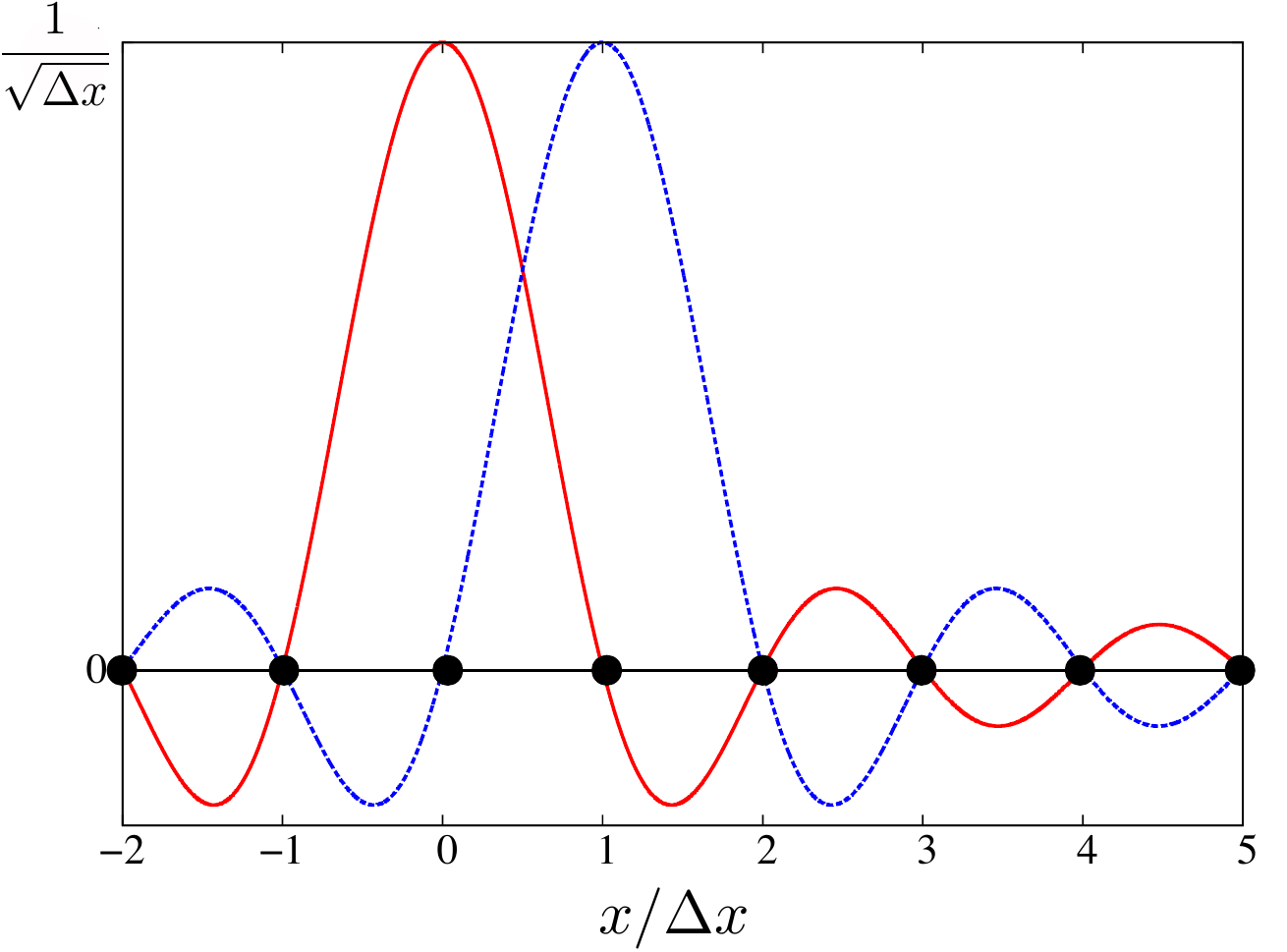}}
\caption{\label{fig:SincFunc} 
\emph{{Sinc DVR basis functions}} The sinc DVR basis functions $\langle x|\Delta_0\rangle$ (red solid) and $\langle x|\Delta_1\rangle$ (blue dashed) are nonzero at their centering grid point and vanish at all other grid points.  The black dots denote the equally spaced grid points.}
\end{figure}

The sinc DVR~\cite{Colbert_Miller_92} is obtained by letting the domain be the real line $\mathcal{D}=\left(-\infty,\infty\right)$, choosing the projection operator $\mathcal{P}_K$ to project into the subspace of wave functions whose bandwidth is limited by a momentum $K$,
\begin{align}
\mathcal{P}_K&=\int_{-K}^{K}dk |k\rangle\langle k|\, ,
\end{align}
where $\langle x|k\rangle=e^{ikx}/\sqrt{2\pi}$ are plane waves with a delta-function normalization, and $N$ equally spaced grid points $x_n=n\pi/K$.  The projection $\mathcal{P}_K$ can also be viewed as projection onto the subspace of the Hilbert space spanned by non-interacting states with energy less than $E_K=\hbar^2K^2/2M$, $M$ being the mass.  It can be verified by direct substitution that delta functions projected into $\mathcal{S}_K$ are in fact sinc functions
\begin{align}
\label{eq:sincfuncs}\langle x|\Delta_n\rangle&=\frac{1}{\sqrt{\Delta x}} \sinc\left(\pi \left(x-x_n\right)/\Delta x\right)\, ,
\end{align}
where $\Delta x=\pi/K$ is the grid spacing, and that these functions are nonzero at a single grid point and vanish on all others, thus satisfying the orthonormality property of a DVR, see Fig.~\ref{fig:SincFunc}.  As will be discussed further in the section of convergence, the momentum-space cutoff $K$ is related to the real-space cutoff $\Delta x$, and so the convergence behavior of DVR methods can be interpreted either in real or momentum space.

One notes from Eq.~\eqref{eq:sincfuncs} that the sinc DVR functions define a quadrature rule~\cite{press1993} with abscissas $\left\{x_n\right\}$ and uniform weights $w_n=\Delta x$ such that the overlap between two such states is \emph{exact}:
\begin{align}
 &\int dx \langle \Delta_n|x\rangle\langle x|\Delta_{n'}\rangle\\
\nonumber &=\sum_{i=-\infty}^{\infty}  \frac{\Delta x}{\Delta x}\sinc\left(\pi\left(i-n\right)\right)\sinc\left(\pi\left(i-n'\right)\right)=\delta_{n,n'}\, .
\end{align}
This remarkable property is also the underpinning for the Nyquist-Shannon sampling theorem, which states that any band-limited function can be completely determined from a sequence of equally spaced samples of the function~\cite{press1993}.  Noting that $\pi/\Delta x$ is the spatial Nyquist frequency for a function with bandwidth $K$, the representation of a band-limited function in DVR basis functions is nothing but Shannon's interpolation formula.

More useful than the fact that we can integrate products of basis functions (and hence, products of any two functions in $\mathcal{S}_K$) exactly with the given quadrature is the fact that integration by this quadrature produces exponentially accurate results for potentials $V\left(x\right)$ which are smooth and slowly varying (i.e.~well-approximated within $\mathcal{S}_K$~\footnote{\label{footnote:OON}If $V(x)\langle x |\Delta_n\rangle\in \mathcal{S}_K$, then there is clearly no error in integrating with the DVR quadrature.  However, the components of $V(x)\langle x |\Delta_n\rangle$ which lie outside of $\mathcal{S}_K$ are typically $\mathcal{O}(1/N)$ and this error is concentrated in the basis elements near the boundary, $n\simeq N$.  For wave functions well-converged with a basis size $N$, the weight on the basis functions with $n$ near $N$ is exponentially small, and hence we recover exponential accuracy~\cite{Littlejohn_Cargo_General}.}).  Here, exponential accuracy refers to convergence with $N$.  The exponential accuracy of the DVR quadrature leads to the matrix elements of the potential within the DVR basis states, $\langle \Delta_n|V\left(x\right)|\Delta_{n'}\rangle\equiv V_{nn'}\approx V\left(x_n\right)\delta_{nn'}$.  As another example of the use of the DVR quadrature, the derivative of a band-limited function is another band-limited function of the same bandwidth, and so the quadrature above provides an exponentially accurate representation of differential operators acting on functions in $\mathcal{S}_K$ via
\begin{align}
\langle \Delta_n |\frac{\partial^k}{\partial x^k}|\Delta_{n'}\rangle&=\frac{\partial^k}{\partial x^k}\sinc\left(\pi x \right)\Big|_{x=n-n'}\, .
\end{align}
For example, this gives that the derivative of a function expressed in the DVR basis
\begin{align}
\psi\left(x\right)&=\sum_n \psi_n\langle x|\Delta_n\rangle
\end{align}
is given by
\begin{align}
\frac{d \psi\left(x\right)}{d x}&=\frac{1}{\Delta x}\sum_n \sum_{l\ne n}\frac{\left(-1\right)^{n-l}}{n-l} \psi_l \langle x|\Delta_n\rangle\, ,
\end{align}
and that a representation of the kinetic energy operator in the DVR basis is given by
\begin{align}
\nonumber T_{nn'}&\equiv \langle \Delta_n|-\frac{\hbar^2}{2M}\frac{\partial^2}{\partial x^2}|\Delta_{n'}\rangle\\
&=\frac{\hbar^2}{2M\Delta x^2}\left\{\begin{array}{c} \frac{\pi^2}{3}\, ,\;\;\;\mbox{$n=n'$}\\ \frac{2\left(-1\right)^{n-n'}}{\left(n-n'\right)^2}\, ,\;\;\; \mbox{otherwise}\end{array}\right. \, .
\end{align}
One can show~\cite{Colbert_Miller_92} that these matrix elements also correspond to the limit of an $N^{\mathrm{th}}$ order finite-difference approximation to the second derivative as $N\to \infty$.  This provides a useful alternative viewpoint of the sinc DVR as a limiting case of grid-based finite difference methods.  While in one dimension the resulting Hamiltonian matrix is no longer sparse due to the fact that the kinetic energy operator is long-ranged in the grid space, in higher (tensor product space) dimensions the Hamiltonian is sparse in the sense that it is block diagonal along each dimension.  In addition, while a finite-order finite differencing scheme would produce a more sparse representation of the kinetic energy than the DVR, such methods are only polynomially convergent.  Hence, the loss in sparsity of the DVR compared to finite-order finite differencing methods is more than compensated by the increase in accuracy.

A particularly nice corollary of the fact that expectation values of functions well-approximated within $\mathcal{S}_K$ are exponentially accurate, which seems not to have been recognized yet in the literature, is that the DVR representation gives an exponentially convergent method for the evaluation of integrals of products of eigenfunctions and possibly also their derivatives.  Such integrals appear in the evaluation of pseudopotential matrix elements Eq.~\eqref{eq:swavepp}-\eqref{eq:pwavepp}, and their proper evaluation is key to quantitative connections between few and many-body physics.  A demonstration of the convergence of interaction matrix elements is given in Sec.~\ref{sec:Convergence}.

For potentials with mirror symmetry, $V\left(-x\right)=V\left(x\right)$, we can divide the space of wave functions into those with even or odd parity about $x=0$.  Correspondingly, we can use the parity-adapted sinc DVR basis sets
\begin{align}
|\Delta_n^+\rangle&=\frac{1}{\sqrt{2\left(1+\delta_{n,-n}\right)}}\left(|\Delta_n\rangle+|\Delta_{-n}\rangle\right)\, ,\;\; n\in\left[0,N\right]\\
|\Delta_n^-\rangle&=\frac{1}{\sqrt{2}}\left(|\Delta_n\rangle-|\Delta_{-n}\rangle\right)\, ,\;\; n\in\left[1,N\right]\, ,
\end{align}
which form complete bases for the even and odd functions in $\mathcal{S}_K$, respectively.  The kinetic matrix elements in these bases are
\begin{align}
T^{+}_{nn'}&=\frac{\hbar^2}{2M \Delta x^2}\left\{\begin{array}{c} \frac{\pi^2}{3}-\frac{2\left(-1\right)^{n+n'}}{\left(n+n'\right)^2}\, ,\;\; n=n'\\ \end{array}\right.\, ,\\
T^{-}_{nn'}&=\frac{\hbar^2}{2M \Delta x^2}\left\{\begin{array}{c} \frac{\pi^2}{3}-\frac{2\left(-1\right)^{n+n'}}{\left(n+n'\right)^2}\, ,\;\; n=n'\\ \frac{2\left(-1\right)^{n-n'}}{\left(n-n'\right)^2}-\frac{2\left(-1\right)^{n+n'}}{\left(n+n'\right)^2}\, ,\;\; n\ne n' \end{array}\right. \,,
\end{align}
and the potential matrix elements remain unchanged, i.e. $V^{\pm}_{n n'}=V_{nn'}\approx V\left(x_n\right)\delta_{nn'}$.

\subsubsection{Bessel DVR}
\label{sec:BDVR}
\begin{figure}[tbp]
\centerline{\includegraphics[width=0.8 \columnwidth]{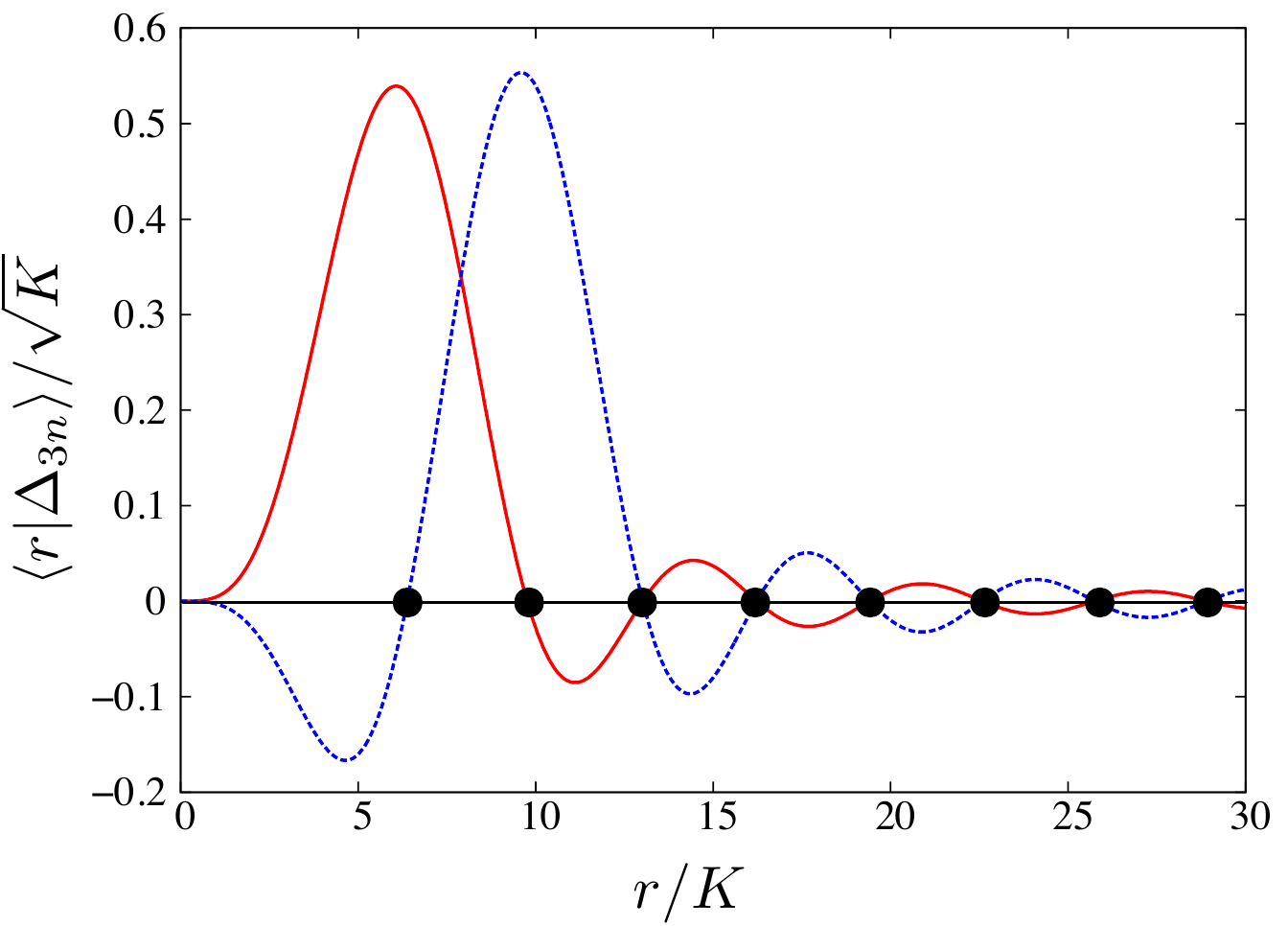}}
\caption{\label{fig:BessFunc} 
\emph{{Bessel DVR basis functions}} The Bessel DVR basis functions $\langle r|\Delta_{3,1}\rangle$ (red solid) and $\langle r|\Delta_{3,2}\rangle$ (blue dashed) are nonzero at their centering grid point and vanish at all other grid points.  Here, the grid (black points) is not uniformly spaced.}
\end{figure}

The Bessel DVR is similar to the above sinc DVR, but uses the free particle wave functions relevant for a radial coordinate.  Namely, we consider the functions $\phi_m\left(r\right)=\sqrt{r}\psi_m\left(r\right)$ in 2D, where $\psi_m\left(r\right)$ are the solutions of the radial Schr\"odinger equation for angular momentum $m\in\mathbb{Z}$.  The radial functions $\phi_m\left(r\right)$ satisfy the equation
\begin{align}
\frac{\hbar^2}{2M}\left[-\frac{d^2\phi\left(r\right)}{dr^2}+\frac{m^2-1/4}{r^2}\phi\left(r\right)\right]+V\left(r\right)\phi\left(r\right)&=E\phi\left(r\right)\, ,
\end{align}
and so behave as $r^{m+1/2}$ near the origin.  For free particles, $V\left(r\right)=0$, the solutions are Bessel functions: $\phi_{mk}\left(r\right)=\sqrt{k r}J_{m}\left(k r\right)\equiv \langle r|k m\rangle$, where $k$ denotes that this is the solution with energy $\hbar^2k^2/2m$.  As with the plane wave states, these free radial states obey a delta-function normalization.  We now obtain a Bessel DVR on the radial domain $\mathcal{D}=\left[0,\infty\right)$ by choosing the projector $\mathcal{P}_{Km}$ to project into the space of free wave functions with energy less than $E_K=\hbar^2K^2/2M$ which vanish as $r^{m+1/2}$ at the origin, 
\begin{align}
\mathcal{P}_{Km}&=\int_0^K dk |k m\rangle\langle k m|\, ,
\end{align}
and letting the grid points correspond to the zeroes  of the function $\langle r| K m\rangle$, the free particle wave function evaluated at the momentum cutoff $K$.  We note that this grid is not uniformly spaced, and depends upon the angular momentum $m$.  Denoting the $n^{\mathrm{th}}$ zero of $J_{m}\left(x\right)$ as $z_{mn}$ ($n=1,\dots,\infty$), the DVR basis functions are
\begin{align}
\langle r|\Delta_{m, n}\rangle&=\left(-1\right)^{n}\frac{K z_{m n}\sqrt{2 r}}{K^2 r^2-z_{mn}^2}J_{m}\left(K r\right)\, ,
\end{align}
and the grid points are $z_{mn}/K$.  Examples of these functions for $m=3$ are shown in Fig.~\ref{fig:BessFunc}.

As was the case with the sinc DVR functions, the Bessel DVR functions also define a quadrature which is exponentially convergent for functions in $\mathcal{S}_{K}$.  One may be concerned about the accuracy of the kinetic energy operator evaluated with DVR quadrature, as the centrifugal potential is singular and hence not within $\mathcal{S}_K$.  However, our DVR basis functions are constructed out of the eigenfunctions of the kinetic energy, and so the DVR basis also represents the full kinetic energy operator with exponential accuracy.  Stated differently, the singular contributions from the derivative operators and the centrifugal potential cancel within our basis, and the remainder is well-represented in $\mathcal{S}_K$.  The matrix elements are given as~\cite{Littlejohn_Cargo_Bessel}
\begin{align}
\nonumber &\langle \Delta_{m, n}|\frac{\hbar^2}{2M}\left[-\frac{d^2}{dr^2}+\frac{m^2-1/4}{r^2}\right]|\Delta_{m, n'}\rangle\\
&=\frac{\hbar^2 K^2}{2M}\left\{\begin{array}{c}\frac{1}{3}\left(1+\frac{2\left(m^2-1\right)}{z_{m n}^2}\right)\, ,\;\; n=n'\\ \left(-1\right)^{n-n'}\frac{8 z_{mn}z_{mn'}}{\left(z_{mn}^2-z_{mn'}^2\right)^2}\, ,\;\;n\ne n'\end{array}\right.\, .
\end{align}

The DVR quadrature associated with the Bessel DVR is less useful than that associated with the sinc DVR due to the fact that the grid points are different for different values of the angular momentum $m$.  However, overlaps between any two functions which have the same value of $m$ still can be evaluated with exponential accuracy by this means.  The evaluation of interaction matrix elements for radially symmetric potentials is discussed in more detail in Appendix~\ref{sec:RInteractions}.

\subsubsection{Convergence}
\label{sec:Convergence}

In both the sinc DVR and Bessel DVR there exist two convergence parameters.  The first is the number of grid points $N$, and the second is the finite domain size of the DVR grid, which can either be taken as a cutoff in real space or momentum space.  While the domain of the DVR grid points is finite, we stress that the DVR basis functions themselves still exist on infinite or half-infinite domains.  The relation between the real space cutoff $R$ and the momentum space cutoff $K$ is $R=N\pi /K$ for the sinc DVR and $R=z_{mN}/K$ for the Bessel DVR.  In this section, we demonstrate the exponential convergence of the sinc DVR method with $N$ and $R$ using a 1D Gaussian well.  The convergence behavior of the Bessel DVR method is similar, and we do not demonstrate it here.

\begin{figure}[tbp]
\centerline{\includegraphics[width=0.8 \columnwidth]{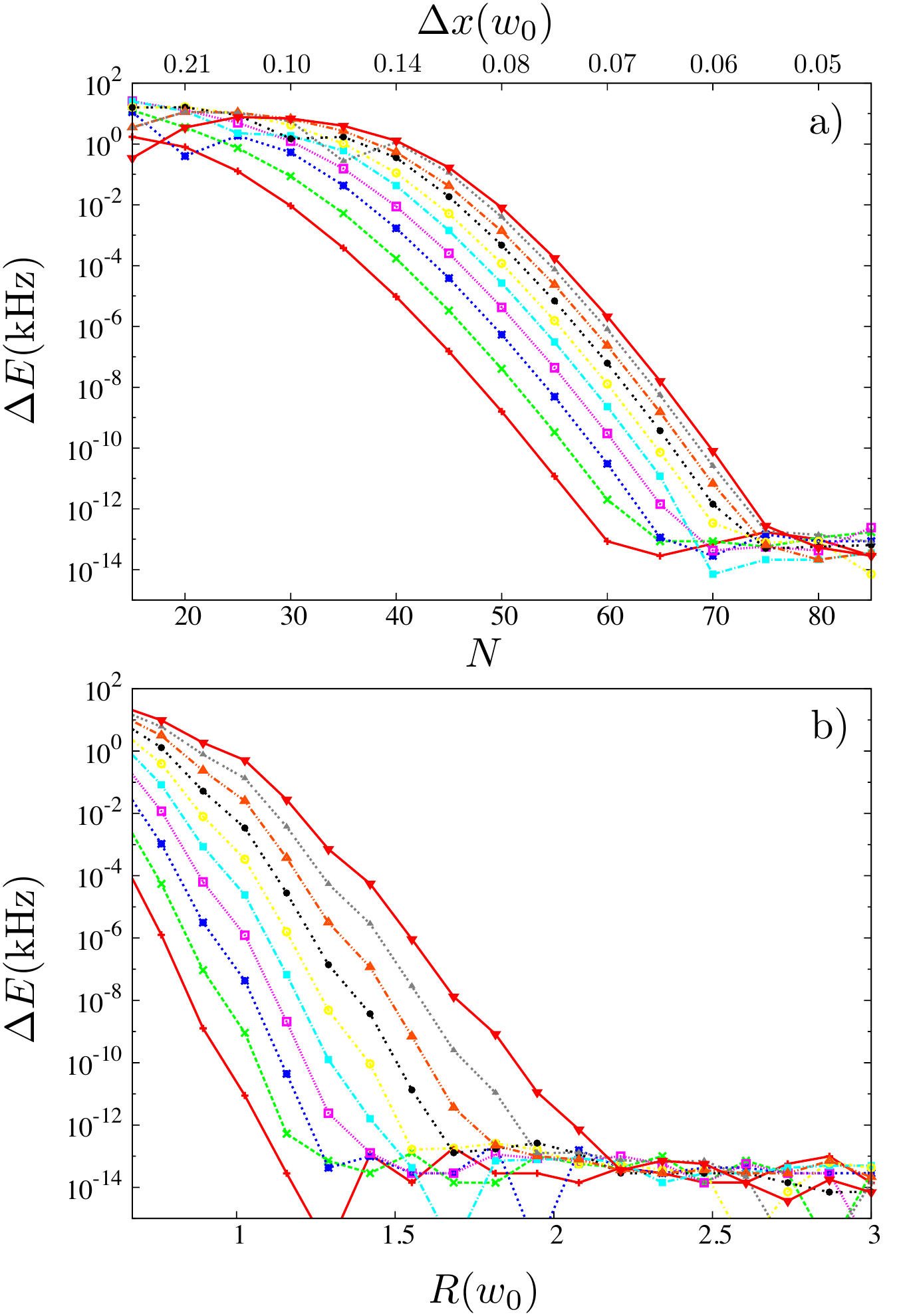}}
\caption{\label{fig:EnergyConvergence} 
\emph{{Convergence of DVR energies with number of DVR grid points and domain size}.} (a) Convergence in the number of DVR grid points (equivalently, the grid spacing $\Delta x$) is demonstrated for the first 10 eigenstates of a Gaussian well at fixed $R=3w_0$.  The points are differences in energies for neighboring $N$, as described in the main text.  Curves from bottom to top are increasing in energy.  (b) The analogous plot to (a) at fixed $\Delta x=0.05$ and varying $R$, the DVR grid domain cutoff.  The parameters used are: depth $V=96\;$kHz, waist $w_0=707\;$nm, and the mass of $^{87}$Rb.}
\end{figure}

In Fig.~\ref{fig:EnergyConvergence} we demonstrate the exponential convergence of the energies for a 1D Gaussian potential well
\begin{align}
V\left(x\right)&=-Ve^{-2x^2/w_0^2}\, ,
\end{align}
where we take $V= 96\;$kHz\footnote{Energies quoted in Hz always mean units of $h$Hz.  We suppress the $h$ for brevity.}, $w_0=707\;$nm, and the mass of $^{87}$Rb.  This convergence is monitored by increasing a convergence parameter $\eta$ in discrete steps $\delta$, and then observing an exponential decrease in $\Delta E(\eta)\equiv E(\eta+\delta)-E(\eta)$.  Panel (a) demonstrates exponential convergence with the number of DVR grid points $N$ at fixed domain size $R$, which can also be stated as convergence in the grid spacing $\Delta x$.  The different curves denote different eigenstates, with lower curves corresponding to lower energy eigenstates.  In panel (b) we show exponential convergence in the domain size $R$ at fixed $\Delta x$.  In this case, convergence of the first 10 eigenenergies to machine precision can be achieved with only 60 grid points (in the parity-adapted DVR basis set).

\begin{figure}[tbp]
\centerline{\includegraphics[width=0.8 \columnwidth]{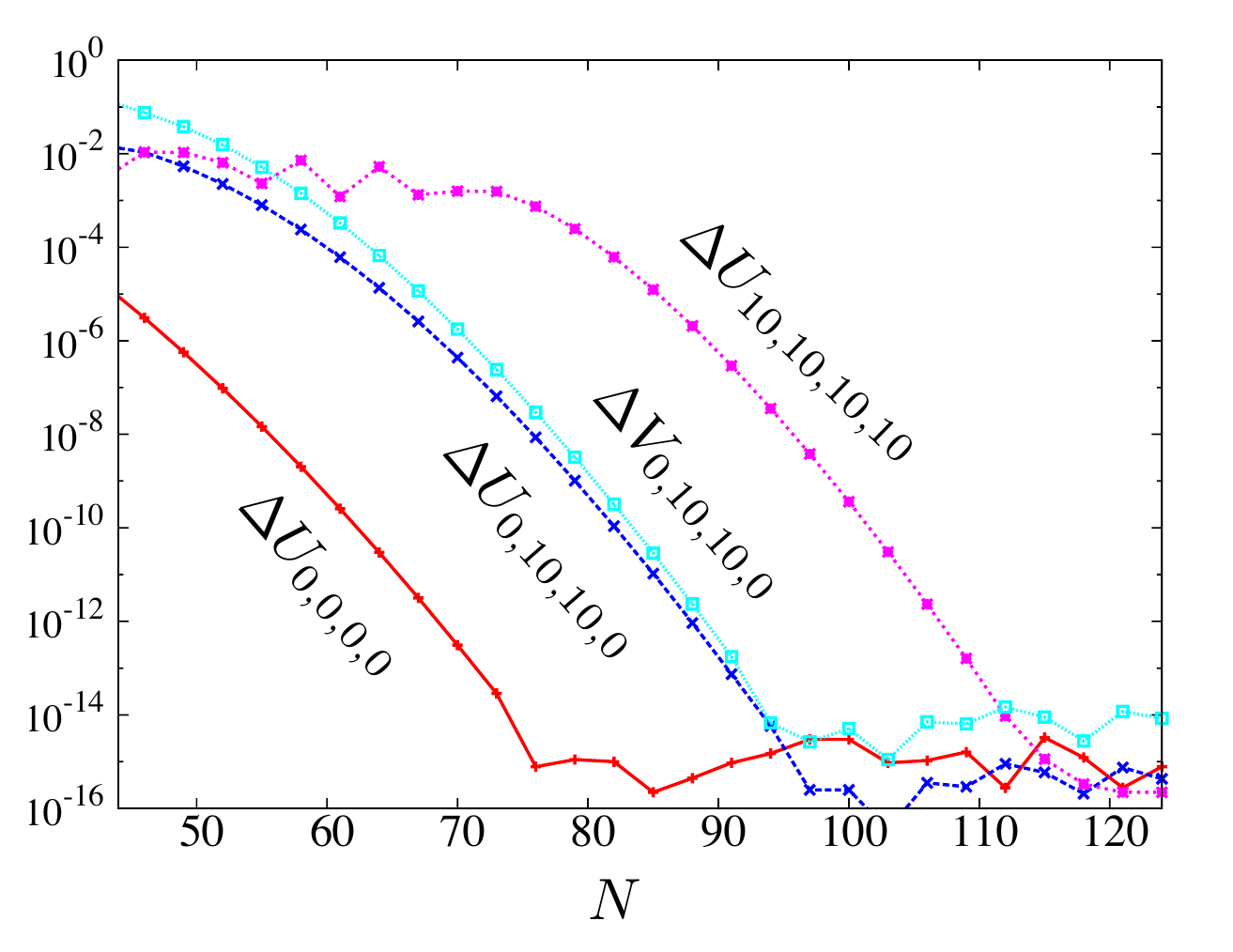}}
\caption{\label{fig:InteractionConvergence} 
\emph{{Convergence of $s$- and $p$-wave interaction matrix elements in the DVR basis size.}} The convergence of the one-dimensional analogs of Eq.~\eqref{eq:swavepp} and Eq.~\eqref{eq:pwavepp} in oscillator units is given by considering the differences with neighboring convergence parameters.  These integrals show a similar rate of convergence to the energies in Fig.~\ref{fig:EnergyConvergence}(a), though the exponential convergence sets in at a larger value of $N$.  The trap parameters are the same as those used in Fig.~\ref{fig:EnergyConvergence}.}
\end{figure}

In Fig.~\ref{fig:InteractionConvergence}, we demonstrate the convergence of the one-dimensional $s$- and $p$-wave interaction integrals, Eq.~\eqref{eq:swavepp} and Eq.~\eqref{eq:pwavepp}, for the same parameters as Fig.~\ref{fig:EnergyConvergence}.  Namely, we plot the differences in the dimensionless parameters $U_{0,0,0,0}a_{\mathrm{ho}}$, $U_{10,10,10,10}a_{\mathrm{ho}}$, $U_{0,10,10,0}a_{\mathrm{ho}}$, and $V_{0,10,10,0}a_{\mathrm{ho}}^{3}$ with neighboring convergence parameters, where $a_{\mathrm{ho}}=\sqrt{\hbar/m\omega}$ is the harmonic oscillator length corresponding to the trap curvature.  As with the energies, we use differences between neighboring convergence parameters, e.g. $\Delta U_{0,0,0,0}(N)=(U_{0,0,0,0}(N+\delta N)-U_{0,0,0,0}(N))a_{\mathrm{ho}}$ to gauge convergence.  We have also checked the convergence of the interaction matrix elements for the harmonic oscillator, where analytic results are available, and found similar convergence.  The rate of convergence of the matrix elements is akin to that of the energies.  However, exponential convergence sets in at a larger value of $N$ for interactions compared to the convergence of the energy.  Convergence in $R$ has a similar qualitative behavior.  Due to the non-uniform grid of the Bessel DVR, computation of the interaction parameters for radial functions is more involved, and is discussed in Appendix~\ref{sec:RInteractions}.

In order to converge results, we find it is useful to first converge in $\Delta x$ and then increase $R$ until convergence.  Convergence with $\Delta x$ can be ascertained by convergence of the energy at fixed $R$, and then convergence with $R$ can be judged by requiring that the weight on the outermost DVR basis functions becomes on the order of machine precision.  In higher-dimensional scenarios, visualization of the convergence with $R$ can be greatly assisted by plotting the nearest separable state.  A precise definition of the nearest separable state and a procedure for calculating it are discussed in Sec.~\ref{sec:NearestSeparable}.

\subsection{Optimizations: Using Sparsity and Variational and Quasi-adiabatic methods}
\label{sec:QAB}

A nice feature of DVRs in dimensions higher than one are that the resulting multi-dimensional Hamiltonian descriptions are sparse.  In particular, for a 3D problem in Cartesian coordinates, the Hamiltonian may be written in a product sinc DVR basis as
\begin{align}
\nonumber &\langle \Delta_x\Delta_y\Delta_z|\hat{H}|\Delta_x'\Delta_y'\Delta_z'\rangle=\delta_{y,y'}\delta_{z,z'}T_{xx'}+\delta_{x,x'}\delta_{z,z'}T_{yy'}\\
&+\delta_{x,x'}\delta_{y,y'}T_{zz'}+\delta_{x,x'}\delta_{y,y'}\delta_{z,z'}V\left(x,y,z\right)\, .
\end{align}
Hence, the Hamiltonian can be applied to a vector describing the state in the DVR basis using $\mathcal{O}\left(N_x^2N_yN_z+N_xN_y^2N_z+N_xN_yN_z^2\right)$ operations, far fewer than the $\mathcal{O}\left(N_x^2N_y^2N_z^2\right)$ operations required for a dense matrix description.  This enables the use of sparse matrix methods requiring only matrix-vector products, such as the Lanczos algorithm~\cite{Golub_VanLoan_96}, to find extremal eigenstates.  Utilizing the sparsity of the representation is also key for efficient simulation of the dynamics of particles in time-varying potentials, as it enables the use of Krylov subspace approximations to the matrix exponential~\cite{Saad_92,Hochbruck_Lubich_97}.  Such dynamical simulations are useful, for example, for determining the degree of adiabaticity when optical tweezer wells are dynamically repositioned.

For lattice problems, using a set of basis functions which has the same translational symmetry as the lattice often leads to significant computational gains.  The DVR basis functions given above are defined on infinite or semi-infinite spaces, and so are inappropriate for expanding a function defined on a finite, periodic space.  For 1D periodic potentials, expansion in terms of a plane wave basis is very efficient and accurate.  Hence, it is natural to combine DVR methods for the transverse potential with plane-wave methods along the periodic direction.  In what follows, we combine the two by expanding the full Hamiltonian in a basis consisting of products of transverse ($r$) and lattice ($z$) degrees of freedom.  The number of functions we use for the expansion along, e.g.~, direction $r$ will be denoted $a_r$, and is called the \emph{variational dimension}.

\begin{figure}[tbp]
\centerline{\includegraphics[width=0.8 \columnwidth]{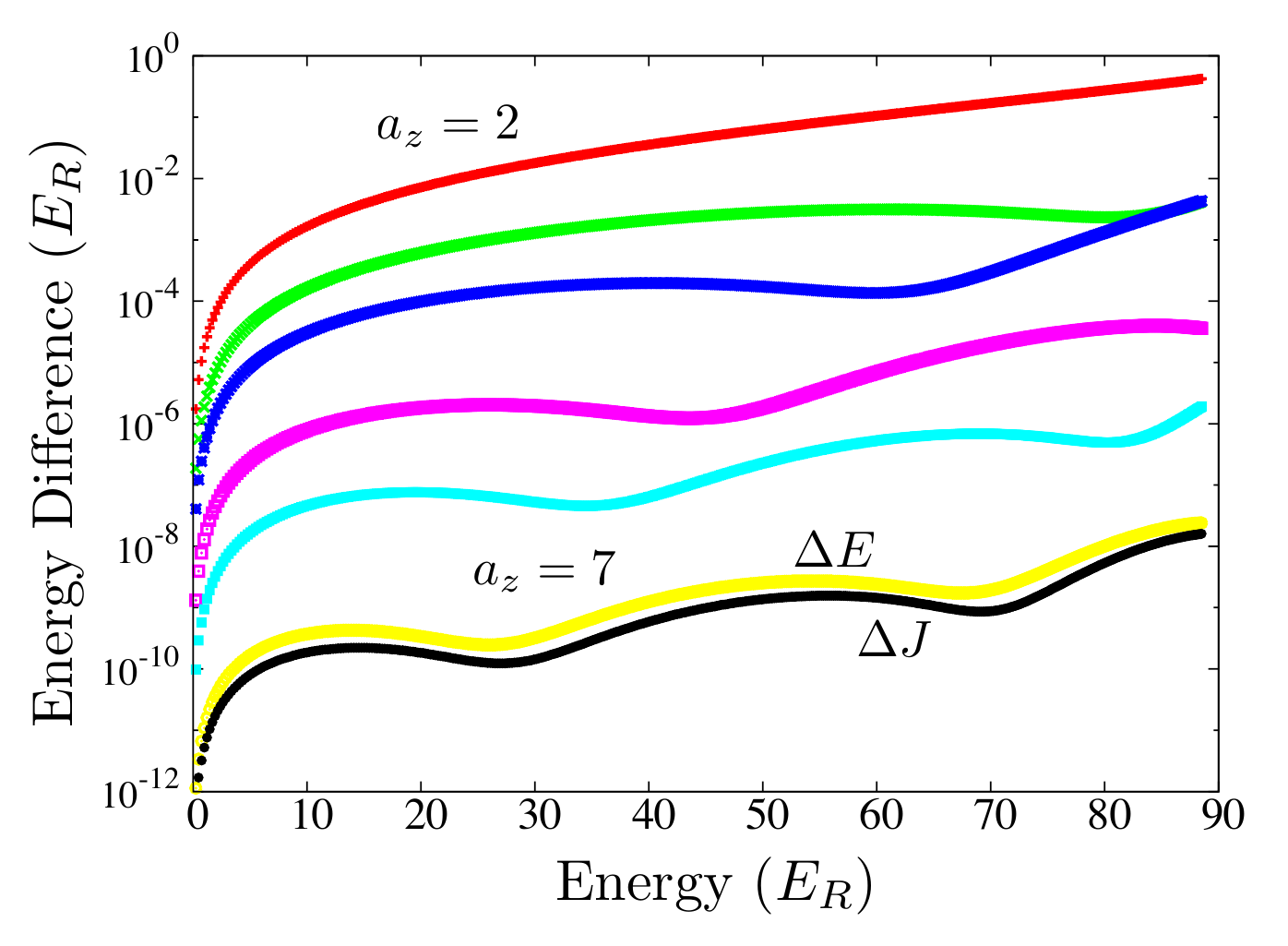}}
\caption{\label{fig:VariationalConvergence} 
\emph{{Convergence of the $m=0$ energies and tunneling amplitudes of the potential Eq.~\eqref{eq:VlattX} with the variational basis size $a_z$.}} The difference in energies adding another lattice band $\Delta E\equiv E(a_z+1)-E(a_z)$ as a function of eigenstate energy are shown for numbers of bands $a_z=2,\dots,7$, with lower curves corresponding to larger numbers of bands.  The bottom curve is $\Delta J$ for $a_z=7$, showing that the tunneling amplitudes are display similar convergence behavior to the energy.  The trap parameters are $w_0=30\;\mu$m, $V_{\mathrm{const}}=88\,E_R$, and $V=12\,E_R$.}
\end{figure}

To see how the combination of DVR methods with other basis sets is facilitated in this situation, consider the lattice potential Eq.~\eqref{eq:Vlatt}.  We can write this potential as $V_{\mathrm{latt}}V_g\left(r\right)V_z\left(z\right)$, where $V_g\left(r\right)=-\exp(-\frac{2r^2}{w_0^2})$ is the Gaussian radial potential and $V_z\left(z\right)=\cos^2\left(k z\right)$ is the lattice corrugation.  This multiplicative separability of the potential provides a natural variational basis with which to expand the full coupled 3D problem, namely products of eigenfunctions of $V_{\mathrm{latt}}V_g\left(r\right)$ with eigenfunctions of $V_{\mathrm{latt}}V_z\left(z\right)$.  Let us denote the eigenfunctions of $V_g(r)$ with angular momentum $m$ as $\mathcal{R}_{m,n_r}\left(r\right)$ and the eigenfunctions of $V_z(z)$ with quasimomentum $q$ as ${\varphi}_{q,n_z}\left(z\right)$, where $n_r$ and $n_z$ are mode and band indices.  Then, the matrix elements of the full Hamiltonian in the product basis $\langle\mathbf{r}|n_rn_z\rangle=e^{i m \phi}\mathcal{R}_{m,n_r}\left(r\right){\varphi}_{q,n_z}\left(z\right)/\sqrt{2\pi}$ are
\begin{align}
\nonumber \langle n_r'n_z'|\hat{H}|n_rn_z\rangle=&T^{\left(r\right)}_{n_r'n_r}\delta_{n_zn_z'}+T^{\left(z\right)}_{n_z'n_z}\delta_{n_rn_r'}\\
&+V_{\mathrm{latt}}V^g_{n_rn_r'}V^z_{n_zn_z'}\, ,
\end{align}
where $O_{\mu \mu'}=\langle \mu |\hat{O}|\mu'\rangle$, $\hat{T}^{\left(r\right)}$ is the radial and azimuthal kinetic energy operator, and $\hat{T}^{\left(z\right)}$ is the $z$ kinetic energy operator.  If we use the $a_r\ll N_r$ lowest-energy states $|n_r\rangle$ to expand the full coupled problem, where we will call $a_r$ the variational dimension and $N_r$ is the DVR basis size, $V^g_{n_rn_r'}$ can be efficiently obtained with $\mathcal{O}\left(a_r^2N_r\right)$ operations within a DVR calculation, as the matrix $V$ is diagonal in the DVR representation.  The kinetic energy matrix elements can then be obtained with $\mathcal{O}\left(a_r^2\right)$ operations as $T_{nn'}=\delta_{nn'}E_n-V_{nn'}$, $E_n$ being the eigenenergy of state $n$.  Restricting to a variational basis results in an eigenvalue problem of linear dimension $a_ra_z$ whose lowest energy solutions converge to the true solutions with $a_r, a_z\to \infty$.  The advantage of this procedure is that the variational dimensions $a_r$ and $a_z$ can be significantly smaller than the ``bare" basis sizes $N_r$ and $N_z$ required to converge the variational basis functions $\mathcal{R}$ and ${\varphi}$.  This procedure can also be applied to the potential Eq.~\eqref{eq:VlattX}.  Here, we take the basis states ${\mathcal{R}}$ to be the eigenvectors of the Hamiltonian with potential $\left(V_{\mathrm{const}}+V\right)V^g\left(r\right)$ and the states ${\varphi}$ to be the eigenvectors of the Hamiltonian with potential $VV^z\left(z\right)$.  The variational Hamiltonian has matrix elements
\begin{align}
\nonumber \langle n_r'n_z'|\hat{H}|n_rn_z\rangle&=T^{\left(r\right)}_{n_r'n_r}\delta_{n_zn_z'}+T^{\left(z\right)}_{n_r'n_r}\delta_{n_zn_z'}\\
&+V^g_{n_rn_r'}\left(V_{\mathrm{const}}\delta_{n_zn_z'}+VV^z_{n_zn_z'}\right)\, .
\end{align}
In principle, one could optimize the depths of the potentials used to compute the functions $\mathcal{R}$ and ${\varphi}$ such that the variational dimensions $a_r$ and $a_z$ are minimal, but in practice using the depths given above works well and we have not explored this possibility.  We demonstrate the convergence of this variational procedure in the number of bands kept $a_z$ with the potential Eq.~\eqref{eq:VlattX} with $w_0=30\;\mu$m, $V_{\mathrm{const}}=88\,E_R$, $V=12\,E_R$, where $E_R=\hbar^2 \pi^2/2m a^2$ is the recoil energy of $^{87}$Sr in a magic lattice of spacing $a=406.72\;$nm.  Fig.~\ref{fig:VariationalConvergence} displays the results, showing rapid convergence in both the energies and tunneling amplitudes.  The transverse variational dimension is $\sim 500$.  For comparison, the DVR basis size is $\sim 2000$, and the number of plane waves used to expand the lattice potential is $\sim 200$.

For optical tweezers, in which the Rayleigh range cannot be neglected, the potential no longer becomes a a product of functions in the transverse and $z$ degrees of freedom.  In this case, one can employ a quasi-adiabatic variational method, in which a set of basis functions are constructed for each axial ($z$) DVR basis state $|\Delta_{z_n}\rangle$ using only the transverse kinetic energy, and then the full problem is diagonalized in this basis.  As an example, consider the single-well optical tweezer potential Eq.~\eqref{eq:sTweez}.  We construct the basis functions $|n_r,z_n\rangle$ as the $a_r$ lowest energy eigenfunctions of $\hat{T}^{\left(r\right)}+V_s\left(r,z_n\right)$ for each DVR grid point $z_n$, and denote the eigenenergies as $E_{n_r,z_n}$.  These energies then form an effective potential for the $z$ degrees of freedom, which obey the Schr\"{o}dinger equation
\begin{align}
\langle n_r, z_n|\hat{H}|n_r',z_n'\rangle&=T^{\left(z\right)}_{z_nz_n'}\delta_{n_r,n_r'}+E_{n_r,z_n}\delta_{n_r,n_r'}\delta_{z_n,z_n'}\, .
\end{align}
The time required for this complete procedure scales as $\mathcal{O}\left(N_zD(N_r)+D\left(a_rN_z\right)\right)$, where $D(N_{\mu})$ is the time required to find the lowest $a_{\mu}$ eigenvectors of a (possibly sparse) matrix of linear dimension $N_{\mu}$.  As with the above, reductions in the computational time required for a given accuracy can be achieved when $a_r\ll N_r$.  

A similar procedure can also be applied to the double-well optical tweezer potential Eq.~\eqref{eq:dTweez}, using the multiplicative separability of the $x$ and $y$ potentials.  Here, we first diagonalize the $x$ kinetic energy together with the potential $V_d\left(x,0,z_n\right)$ at each axial DVR grid point $z_n$, keeping the lowest $a_x$ quasi-adiabatic levels.  Next, we diagonalize the $x$ and $y$ kinetic energies together with the full potential constructed from the product of the matrix elements of $V_d\left(x,0,z_n\right)$ in the $x$ quasi-adiabatic basis and a Gaussian along $y$, extracting the lowest $a_{xy}$ quasi-adiabatic energies.  These energies then form an effective potential which is diagonalized with the $z$ kinetic energy to obtain the full solutions in $\mathcal{O}\left(N_zD(N_x)+N_zD(a_xN_y)+D(a_{xy}N_z)\right)$ time.

\section{Matrix product state representations of non-separable states}
\label{sec:NearestSeparable}

The eigenfunctions of a non-separable potential are themselves not separable, meaning that $\psi\left(\mathbf{r}\right)\ne \psi_x\left(x\right)\psi_y\left(y\right)\psi_z\left(z\right)$ or $\psi\left(\mathbf{r}\right)\ne \psi_{\phi}\left(\phi\right)\psi_r\left(r\right)\psi_z\left(z\right)$ for potentials of cylindrical symmetry.  However, in many cases one would expect that the lowest-energy states of sufficiently deep potentials are ``nearly separable," as the potential becomes nearly harmonic over the extent of the wavefunction.  In this section, we provide several quantitative measures of non-separability for the eigenstates of non-separable potentials, and investigate near-separable approximations of non-separable states using tools borrowed from quantum information theory. 

\subsection{Schmidt form for non-separable states with cylindrical symmetry}
 The simplest case to study non-separability in a 3D potential is to consider a system with cylindrical symmetry. Here, the azimuthal degrees of freedom are separable, leaving only the $r$ and $z$ degrees of freedom coupled so that the wave function may be written
\begin{align}
\psi_{m}\left(\mathbf{r}\right)&=\frac{e^{im\phi}}{\sqrt{2\pi}}\zeta_m\left(r,z\right)\, .
\end{align}
We can find the state nearest to the true state in the 2-norm with a restricted amount of separability (in a sense to be made precise below) by using the Schmidt decomposition~\cite{nielsen_chuang_book_00}.  The Schmidt decomposition states that we can write any state on a product Hilbert space $\mathcal{H}_1\otimes \mathcal{H}_2$ as
\begin{align}
\label{eq:psiSchmidt} |\psi\rangle&=\sum_{\mu=1}^{\chi} \lambda_{\mu}|\phi_{\mu}\rangle|\xi_{\mu}\rangle
\end{align}
where the sets $\left\{|\phi_{\mu}\rangle \right\}$ and $\left\{|\xi_{\mu}\rangle \right\}$ consist of orthonormal states in $\mathcal{H}_1$ and $\mathcal{H}_2$, respectively, and the Schmidt coefficients $\left\{\lambda_{\mu}\right\}$ satisfy $\lambda_{\mu}> 0$, $\sum_{\mu}\lambda_{\mu}^2=1$.  The Schmidt decomposition is unique, up to unitary rotations in subspaces with degenerate Schmidt coefficients.  The dimension $\chi$ is called the Schmidt rank, and $\chi=1$ if and only if the state is separable.  Hence, $\chi$ quantifies the degree of non-separability.  A continuous measure of separability is given by the von Neumann entropy $S=-\sum_{\mu}\lambda^2_{\mu} \log \lambda_{\mu}^2$.  We note that the Schmidt decomposition can be obtained efficiently numerically using the singular value decomposition (SVD) of the coefficient matrix of the wavefunction in a given basis, e.g., a DVR basis.  Further, the Schmidt decomposition is invariant under local unitary transformations, which is to say that Eq.~\eqref{eq:psiSchmidt} is unchanged if we change bases in $\mathcal{H}_1$ or $\mathcal{H}_2$.  This implies that if we have our wavefunction expressed in terms of a variational basis formed of products of $r$ and $z$ functions, as described in Sec.~\ref{sec:QAB}, then we can find the Schmidt decomposition by performing the SVD on the matrix of coefficients in the variational basis.  As the variational basis sizes $a_r, a_z$ are often significantly smaller than the grid sizes $N_r,N_z$, the resulting SVD computation is more efficient.

The state nearest to $|\psi\rangle$ with fixed non-separability, i.e. fixed Schmidt rank $\tilde{\chi}\le \chi$, is obtained by truncating the expansion Eq.~\eqref{eq:psiSchmidt} at $\mu=\tilde{\chi}$.  The difference between this truncated state and the true state is given by the discarded Schmidt coefficients 
\begin{align}
\label{eq:SchmidtError}\left||\tilde{\psi}\rangle-|\psi\rangle\right|^2=\sum_{\alpha=\tilde{\chi}+1}^{\chi}\lambda_{\alpha}^2\equiv \varepsilon_{\tilde{\chi}}\, .
\end{align}
In practice, the state $|\tilde{\psi}\rangle$ needs to be renormalized by setting the Schmidt coefficients $\tilde{\lambda}_{\mu}=\lambda_{\mu}/\sqrt{\sum_{\alpha=1}^{\tilde{\chi}}\lambda_{\alpha}^2}$, $\mu=1,\dots,\tilde{\chi}$.  The resulting correction of $(1-\varepsilon_{\tilde{\chi}})^{-1}$ is inconsequential to the error bound of Eq.~\eqref{eq:SchmidtError} as $\varepsilon_{\tilde{\chi}}\to 0$.  For states with a low degree of non-separability such that the Schmidt rank required to reproduce the state with error $\varepsilon$ in the sense of Eq.~\eqref{eq:SchmidtError}, $\tilde{\chi}$, is much less than the maximum allowed $\chi$, using the Schmidt form drastically reduces the memory required for storing quantum states from $N_rN_z$ to $\tilde{\chi}\left(N_r+N_z\right)$, and also reduces operations acting only on one degree of freedom, say the $z$ degree of freedom, from $N_rN_z^2$ to $\tilde{\chi} N_z^2$.  The savings are especially beneficial in cases where thermal averages over a large number of states are to be performed.  Finally, we note that the nearest separable state is obtained by setting $\tilde{\chi}=1$.  In addition to being useful for visualization and developing intuition, the nearest separable state $|\psi_{\mathrm{sep}}\rangle$ defines an additional measure of non-separability $\mathcal{E}\equiv -\log (|\langle \psi_{\mathrm{sep}}|\psi\rangle|^2)$ which we will call the \emph{geometric non-separability}, in analogy with the geometric entanglement~\cite{PhysRevA.68.042307} motivating its definition.  As opposed to the von Neumann entropy, which is a measure of non-separability for a specific bipartition of the degrees of freedom, the geometric non-separability is a global measure of the non-separability of the full state.  This distinction is most important in the multi-partite case discussed in the next section, where multiple possible bipartitions exist.

\subsection{Matrix product state form for non-separable states in Cartesian coordinates}

The above analysis is directly applicable in the case of a single-well optical tweezer or a 1D optical lattice, but not for a system without cylindrical symmetry, such as the double-well optical tweezer Eq.~\eqref{eq:dTweez}.  Here, the state is most directly represented in Cartesian space, and the $x$, $y$, and $z$ degrees of freedom are all coupled.  The natural generalization of the Schmidt decomposition to multipartite systems is a matrix product state (MPS) decomposition~\cite{Schollwoeck_11} (also called a tensor train decomposition in the mathematics literature~\cite{doi:10.1137/090752286}), which in the present case reads
\begin{align}
\label{eq:CartMPS}\psi\left(\mathbf{r}\right)&=\sum_{\mu_{xy}=1}^{\chi_{xy}}\sum_{\mu_{yz}=1}^{\chi_{yz}}X_{\mu_{xy}}\left(x\right)Y_{\mu_{xy}\mu_{yz}}\left(y\right)Z_{\mu_{yz}}\left(z\right)\, .
\end{align}
Here, $\chi_{xy}$ and $\chi_{yz}$ are the Schmidt ranks corresponding to bipartitions of the state into $x$ and $y\otimes z$ degrees of freedom and $x\otimes y$ and $z$ degrees of freedom, respectively.  The savings in using the MPS format for a non-separable state is even more striking than in the two-partite case: storage is reduced from $N_xN_yN_z$ to $(\chi_{xy}N_x+\chi_{xy}\chi_{yz}N_y+\chi_{yz}N_z)$ and operations on, e.g., the $x$ degree of freedom can be applied in $N_x^2\chi_{xy}$ operations rather than $N_x^2 N_yN_z$.  We note that the representation Eq.~\eqref{eq:CartMPS} is not the only MPS topology which can represent a non-separable state in 3D.  For example, we could have chosen the partition $x-y-z$ or $z-x-y$.  These other permutations will generally have different Schmidt ranks $\chi$, which implies a difference in accuracy for describing the state for a given number of parameters.  Here, we do not make any claims about the optimality of the $x-y-z$ partition of degrees of freedom in Eq.~\eqref{eq:CartMPS}; the optimal partition of degrees of freedom must be determined on a case-by-case basis.

The multi-partite representation Eq.~\eqref{eq:CartMPS} bears many similarities with the two-partite Schmidt decomposition Eq.~\eqref{eq:psiSchmidt}.  For example, both approximations are controlled by the Schmidt ranks of bipartitions, and both representations can be obtained via the SVD (recursively applied at each bipartition, in the case of the MPS form), as is shown explicitly in Appendix~\ref{sec:MPSVari}.  However, there are key differences between the MPS form and the Schmidt form.  An important example is in finding the nearest state with a restricted amount of non-separability.  Compression of an MPS to the optimal MPS with restricted non-separability $\chi$ \emph{at a given bipartition} is achieved by truncating the range of $\mu_{{yz}}$ in Eq.~\eqref{eq:MPSyzdef} or $\mu_{xy}$ in Eq.~\eqref{eq:MPSDone}.  The resulting error in the 2-norm is given by the sum of squares of the discarded singular values, as in Eq.~\eqref{eq:SchmidtError}.  However, finding the nearest MPS in a \emph{global} sense is not as simple as for the two-partite case~\cite{1367-2630-12-2-025008, Wei_Thesis}.  Here, variational algorithms which minimize the distance between the true state and one in a variational manifold with fixed non-separability in an iterative fashion perform well.  In Appendix~\ref{sec:MPSVari}, we present a such a variational algorithm for finding the product state $\tilde{X}(x)\tilde{Y}(y)\tilde{Z}(z)$ nearest a given state expressed in the MPS format.

While the above decompositions are useful for visualization, quantification of non-separability, and storage of states, obtaining such a decomposition via the SVD is as computationally demanding as finding the eigenstates themselves.  However, the fact that the states explored in this work are only weakly non-separable in spite of the fact that they are significantly anharmonic hints that the MPS form given by Eq.~\eqref{eq:CartMPS} could be useful as an ansatz which is variationally optimized, similar to the way MPSs are used in the density matrix renormalization group (DMRG) method of condensed matter physics~\cite{white_92}.  Such an algorithm would find states with a restricted degree of non-separabaility using significantly fewer resources than direct diagonalization.  This prospect is especially promising for multi-particle systems, where failure of the center of mass and relative coordinates to separate in an anharmonic potential makes analysis significantly more difficult~\cite{PhysRevA.84.062710,PhysRevLett.110.203202}.  We leave a detailed description of such an algorithm for future work.

\section{Application to optical tweezer arrays and lattices}
\label{sec:Restuls}

In this section, we apply the above algorithms and measures to two cases of interest.  The first is a double-well optical tweezer of the form Eq.~\eqref{eq:dTweez}, as has been considered in recent experiments at JILA~\cite{Kaufman_Lester_14}.  Here, we focus on the properties of the lowest few states.  The second case we study is that of a 1D optical lattice of the form Eq.~\eqref{eq:VlattX}.  Here, we are interested in the spectrum of states spanning a large fraction of the lattice depth, as is often the case when non-degenerate gases are trapped in such lattices.  The reason we study the generalization Eq.~\eqref{eq:VlattX} over Eq.~\eqref{eq:Vlatt}, with the latter corresponding to the potential used in neutral atom optical clocks~\cite{Martin09082013,Hinkley13092013}, is that the potential Eq.~\eqref{eq:VlattX} enables us to study tunneling in a non-separable lattice with a tunable degree of both non-separability and transverse confinement.

\subsection{Optical tweezers}

As an example of our methodologies applied to arrays of optical tweezers, we consider a double-well optical tweezer of the form Eq.~\eqref{eq:dTweez}, and take parameters similar to the JILA experiment~\cite{Kaufman_Lester_14}, with waist $w_0=707\;$nm, Rayleigh range $z_R=2.17\;\mu$m, and mass and $s$-wave scattering length for $^{87}$Rb in the $|F=2,m_F=2\rangle$ state.  Before we give results for Hubbard parameters and quantitative measures of non-separability, we first detail a procedure to obtain localized Wannier-like functions from the numerical eigenstates of the double-well optical tweezer.

\subsubsection{Construction of Wannier functions}

For inversion-symmetric lattice potentials in 1D, the localization properties of Wannier functions can be analyzed in detail.  In particular, a seminal work of Kohn~\cite{Kohn_59} showed that requiring the phases on the Bloch functions $\psi_{nq}\left(x\right)$ to be such that $\psi_{nq}\left(x\right)$ is a smooth function of the quasimomentum $q$ leads to Wannier functions which are real and have an asymptotically exponential decay away from their centering position.  Apart from this situation, a procedure to obtain so-called maximally localized Wannier functions has been extensively developed which minimizes the second moment of Wannier function position away from its center~\cite{RevModPhys.84.1419}.  Here, we put forth a similar criterion for the generation of localized orbitals from the eigenstates of the non-separable double-well optical tweezer potential Eq.~\eqref{eq:dTweez}.  With a slight abuse of terminology, we will also refer to these localized orbitals as Wannier functions.

In particular, we take as our localization functional the probability to measure a particle in the left well.  Mathematically, we use the functional
\begin{align}
\mathcal{L}\left(\psi_A,\psi_B\right)&=\int dy dz\int_{-\infty}^{x_{\mathrm{cut}}}dx\,\psi_A^{\star}\left(\mathbf{r}\right)\psi_B\left(\mathbf{r}\right)\, ,
\end{align}
where $x_{\mathrm{cut}}$ is some parameterization of what defines the left well, e.g.~the center between the minima, the local maximum, or $x=0$.  Now, given some subset of spatially overlapping eigenstates $\{\psi_{k_1}\left(\mathbf{r}\right),\psi_{k_2}\left(\mathbf{r}\right),\dots,\psi_{k_N}\left(\mathbf{r}\right)\}$ we seek the normalized linear combination $w\left(\mathbf{a},\mathbf{r}\right)=\sum_{i=1}^{N} a_i \psi_{k_i}\left(\mathbf{r}\right)$, $\mathbf{a}\cdot\mathbf{a}=1$ satisfying
\begin{align}
\label{eq:Lmax} \max_{\mathbf{a}}\mathcal{L}\left[w\left(\mathbf{a},\mathbf{r}\right),w\left(\mathbf{a},\mathbf{r}\right)\right]\, .
\end{align}
Alternatively, for functions maximally localized on the right, we take the minimum.  Eq.~\eqref{eq:Lmax} can be written as
\begin{align}
\max_{\mathbf{a}} \mathbf{a}\cdot \mathbb{L}\cdot \mathbf{a}\, ,
\end{align}
where the localization matrix has matrix elements
\begin{align}
\mathbb{L}_{ij}&=\mathcal{L}\left[\psi_{k_i}\left(\mathbf{r}\right),\psi_{k_j}\left(\mathbf{r}\right)\right]\, .
\end{align}
$\mathbb{L}$ defines a symmetric positive semidefinite quadratic form, and so the normalized vector maximizing the localization criterion is the eigenvector corresponding to the largest eigenvalue.  Moreover, the complete set of orthonormal eigenvectors of the localization matrix provides a new basis which is optimal according to our localization criterion.  Strictly speaking, in order to find the optimally localized basis according to our criterion, all states at experimentally relevant energies should be included.  However, for states which are well-separated in energy, the mixing between them incurred in the localization transformation is small and has little effect on the Hubbard parameters.  Hence, one can localize subsets of energetically separated states (with energy differences large compared to the tunneling splitting) individually for greater efficiency.  In the case of a symmetric double well, $\Delta=0$, the closely spaced doublets with even and odd parity along the tunneling direction form a $2\times 2$ localization matrix with diagonal elements $0.5$ and small but non-zero off-diagonal elements.  Hence, the Wannier functions in this case are even and odd superpositions of the even and odd parity functions, which are related to one another by the parity transformation.

\subsubsection{Tunneling and effective bias in an asymmetric double-well optical tweezer}

\begin{figure}[tbp]
\centerline{\includegraphics[width=0.8 \columnwidth]{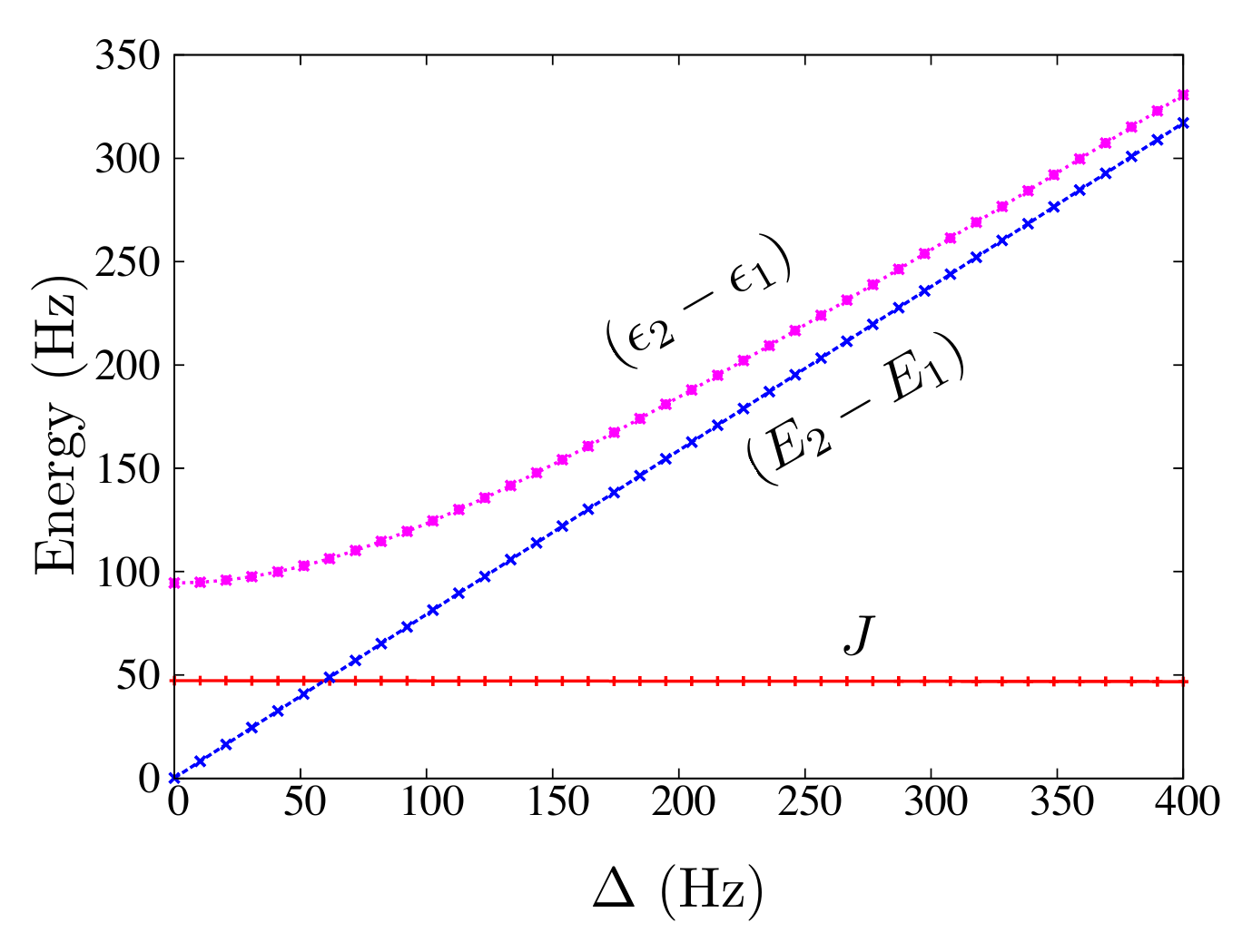}}
\caption{\label{fig:BiasedHubbard} 
\emph{Single-particle Hubbard parameters in an asymmetric double-well optical tweezer}.  The maximal localization procedure described in the text converts the tunneling doublet whose energy splitting $(\epsilon_2-\epsilon_1)$ shown in magenta into a single-band Hubbard model with a nearly-constant tunneling $J$ in red and an effective bias $(E_2-E_1)$ which scales linearly with the applied potential bias.}
\end{figure}

Using the localization prescription given in the last section, we find the tunneling amplitudes $J_{\mu}$ and on-site energies $E_{\mu}$, cf. Eqs.~\eqref{eq:onsiteE}-\eqref{eq:tunn} in the double-well in the basis of maximally localized Wannier functions.  In particular, we have
\begin{align}
E_{\mu}&=\sum_{\nu} (a^{(\mu)}_{\nu})^2 \epsilon_{\nu}\, ,\\
J_{\mu}&=-\sum_{\nu} (a^{(\mu)}_{\nu}a^{(\bar{\mu})}_{\nu}) \epsilon_{\nu}\, ,
\end{align}
where $\mathbf{a}^{(\mu)}$ is the $\mu^{\mathrm{th}}$ eigenvector of the localization matrix, $\epsilon_{\nu}$ is the energy of state $\psi_{k_{\nu}}(\mathbf{r})$, and $\bar{\mu}$ denotes the index of the state connected to $\mu$ by tunneling.

As an example, Fig.~\ref{fig:BiasedHubbard} shows the results for an ${a}=853\;$nm spacing and ${V}=96\;$kHz depth double-well tweezer as a function of the applied bias $\Delta$, focusing on the lowest two states.  These two states are a tunneling doublet when $\Delta=0$, and asymptotically become two localized states with negligible spatial overlap as $\Delta$ becomes large.  The tunneling $J(\Delta)$ in the optimized $\Delta$-dependent Wannier basis changes by less than a percent over the range of biases shown here.  The effective bias $\Delta_{\mathrm{eff}}$, given by the difference in energy $(E_2-E_1)$ between the Wannier functions localized in the two wells, is well-represented by a linear function of the potential bias.  However, the slope of the linear dependence is generally less than one, $\Delta_{\mathrm{eff}}=m \Delta$, $m<1$, due to curvature effects as the potential depth is changed.  In the given example, the best fit gives $m=0.79$.  Also displayed is the difference in energy between the two eigenstates used to construct the Wannier functions, $(\epsilon_2-\epsilon_1)$.  This energy difference fits well to $\sqrt{(2J)^2+\Delta_{\mathrm{eff}}^2}$, validating the single-band Hubbard model description, even when $\Delta\gtrsim J$.

\subsubsection{Tunneling and interaction parameters in a symmetric double-well optical tweezer}

\begin{figure}[tp!]
\centerline{\includegraphics[width=0.8 \columnwidth]{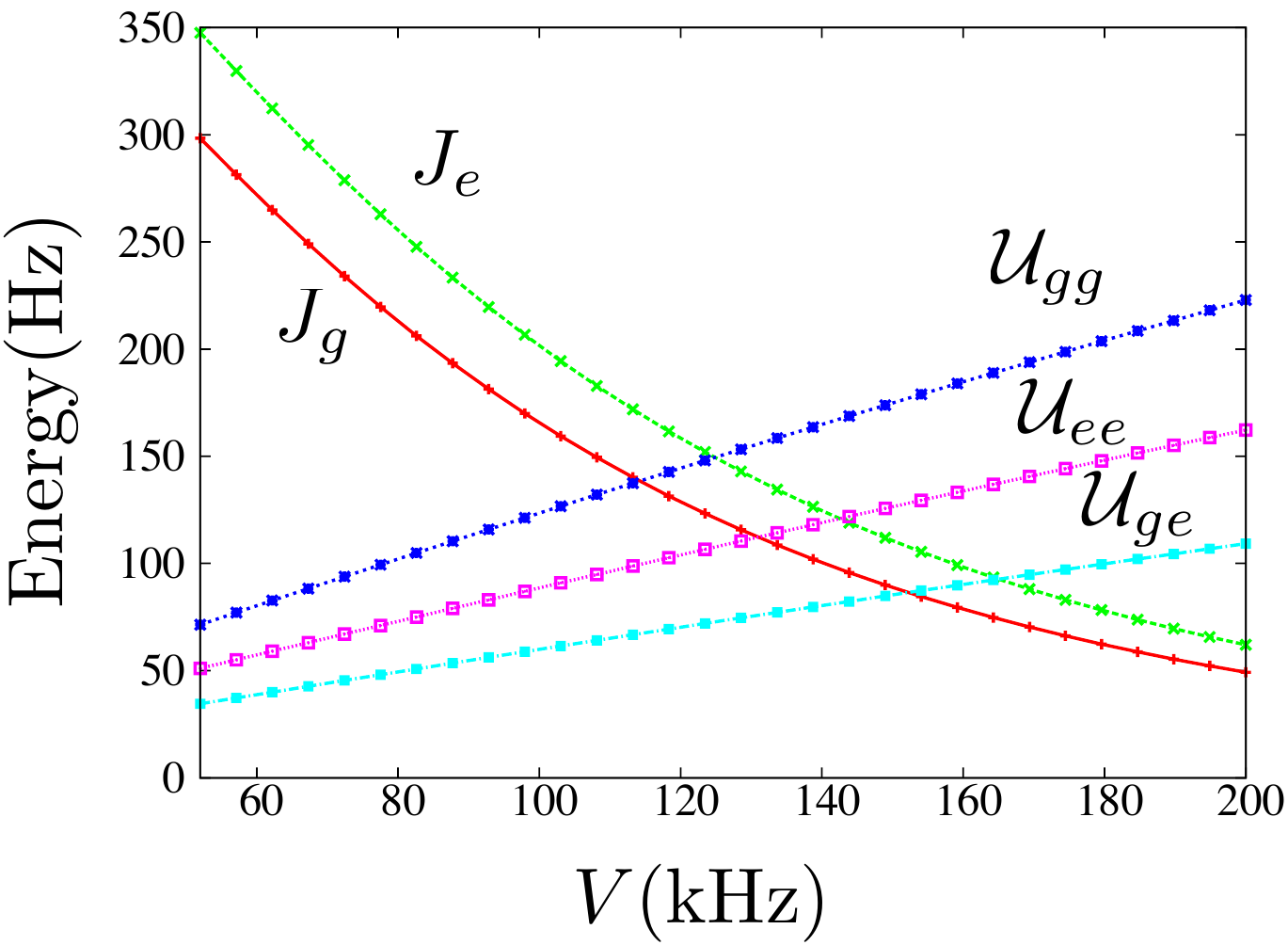}}
\caption{\label{fig:OTHubb} 
\emph{Tunneling and interaction parameters in a symmetric double-well optical tweezer.}  The tunneling and $s$-wave interaction energies for the ground ($g$) and first $z$ excited ($e$) states in a double-well optical tweezer as a function of depth. While the reduction in $U_{ee}$ compared to $U_{gg}$ is an effect of the different spatial extent of the Wannier orbitals in the $g$ and $e$ states that occurs even for separable potentials, the increase of $J_e$ vs.~$J_g$ is solely a manifestation of non-separability.}
\end{figure}

We now turn to the case of of a symmetric double well potential, using as an example the spacing $a=820\;$nm.  In Fig.~\ref{fig:OTHubb}, we show the tunneling and interaction energies for the ground ($g$) state as well as the first axially excited state along $z$ ($e$).  We note that the excitation in the state $e$ is perpendicular to the tunneling direction, and so would not affect tunneling in a separable potential.  The excited state tunneling is roughly 20\% larger than the ground state tunneling as a consequence of non-separability.  To make a quantitative comparison with tunneling in optical lattices, we can approximate our double-well tweezer as an optical lattice with lattice constant $\tilde{a}$ twice the distance from the origin to $\min_{x>0} V\left(x,0,0\right)$ and depth given by the local maximum $\tilde{V}=V\left(0,0,0\right)-V\left(\tilde{a}/2,0,0\right)$.  For the given spacing we consider, $\tilde{a}=642.2\;$nm, the associated energy $E_{\tilde{a}}=\hbar^2\pi^2/2m\tilde{a}^2$ is $1.39\;$kHz, and the effective lattice depth is $\tilde{V}=0.0473V$.  We fit our data in the range $V\in [50\,\mbox{kHz},200\,\mbox{kHz}]$ to the standard formula for tunneling in an optical lattice
\begin{align}
\label{eq:OLfit}J/E_R&=A\left(\frac{V}{E_R}\right)^B \exp\left(-C \sqrt{V/E_R}\right)\, ,
\end{align}
using $E_{\tilde{a}}$ as the recoil energy and $\tilde{V}$ as the lattice depth.  In contrast to the optical lattice, in which the optimal parameters are $A=1.363$, $B=1.057$, and $C=2.117$, we find $A=2.563$, $B=1.217$, and $C=2.281$.  It should be noted that these values also depend on the waist and Gaussian spot spacing.  While the function Eq.~\eqref{eq:OLfit} provides an excellent fit, the optical lattice analogy itself amounts to a $60\%-70\%$ discrepancy over the range of parameters considered, with the tweezer having a larger tunneling for a given effective lattice depth.

Fig.~\ref{fig:OTHubb} also shows the behavior of the $s$-wave interaction energies $\mathcal{U}_{\sigma\sigma'}=\frac{4 \pi \hbar^2 a_s}{M}U_{\sigma \sigma'; \sigma' \sigma}$ as a function of lattice depth.  For an intuitive understanding of the behavior and to quantify the degree of anharmonicity, we will compare with the predictions obtained by using a harmonic approximation for the potential minima.  For harmonic wells, $\mathcal{U}\sim (a_xa_ya_z)^{-1}$, with $a_{\nu}$ the harmonic oscillator length along Cartesian direction $\nu$, and $a_{\nu}\sim V^{-1/4}$, leading to $\mathcal{U}\sim V^{3/4}$.  The best fit for $\mathcal{U}_{gg}$ predicts $\mathcal{U}_{gg}\sim V^{0.85}$, showing a faster rise of interactions with lattice depth than predicted for a harmonic well.  Further, in a harmonic potential, the interactions are related as $\mathcal{U}_{ee}/\mathcal{U}_{gg}=3/4$ and $\mathcal{U}_{eg}/\mathcal{U}_{gg}=1/2$, but due to anharmonicity $\mathcal{U}_{ee}/\mathcal{U}_{gg}\approx 0.714-0.728\,$ and $\mathcal{U}_{eg}/\mathcal{U}_{gg}\approx 0.484-0.49$ for the range of depths in Fig.~\ref{fig:OTHubb}.  

\subsubsection{Separability characteristics of double-well tweezer states}

\begin{figure}[tp!]
\centerline{\includegraphics[width=0.8 \columnwidth]{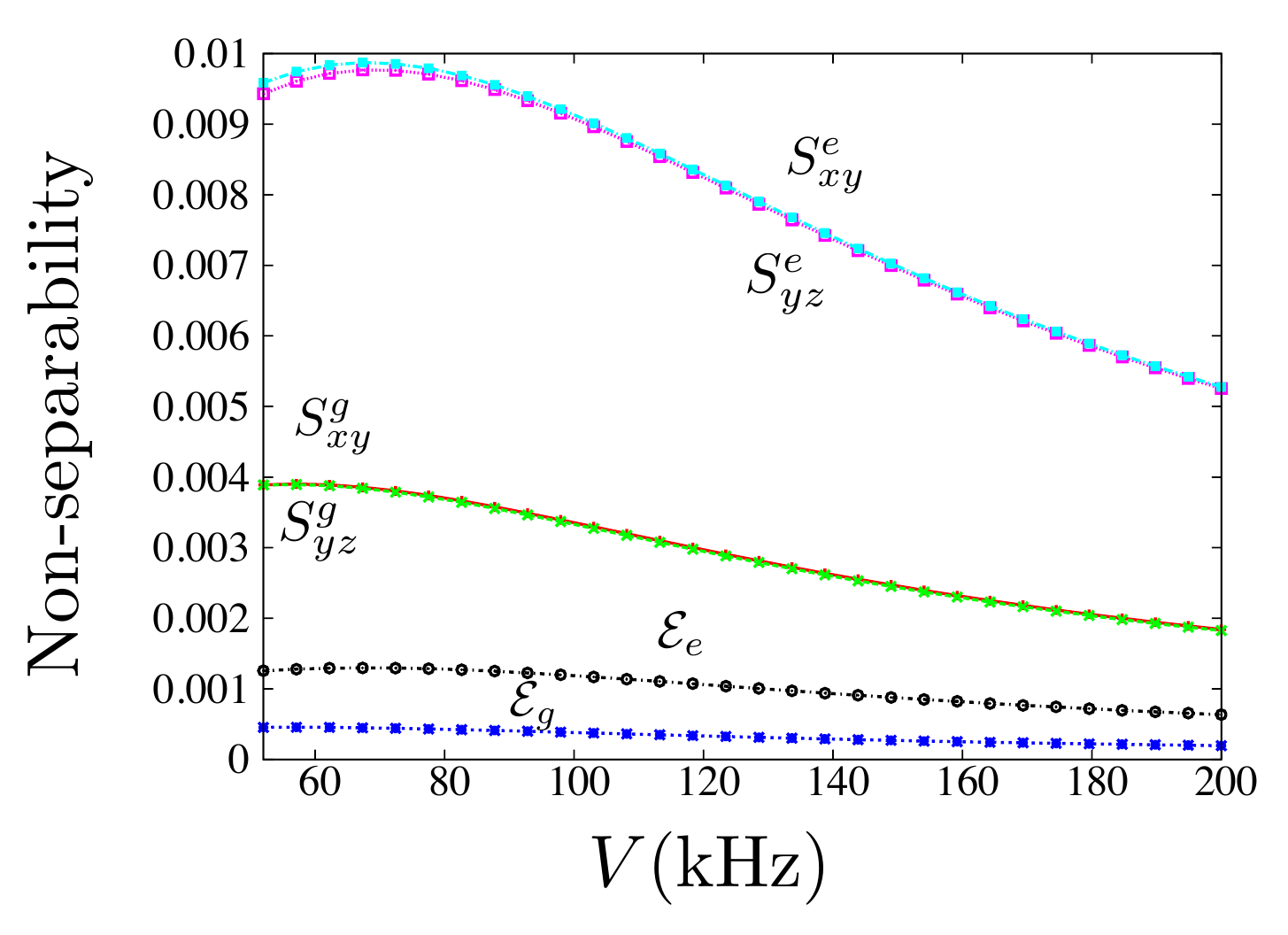}}
\caption{\label{fig:OTNS} 
\emph{Non-separability of states in a symmetric double-well optical tweezer.}  The von Neumann entropy $S$ of $xy$ and $yz$ bipartitions and the geometric non-separability $\mathcal{E}$ for the ground state ($g$) and the first $z$ excited state ($e$).  While both states are near-separable, the excited state is less separable than the ground state.}
\end{figure}

Fig.~\ref{fig:OTNS} shows the von Neumann entropy of non-separability for bipartitions into $x$ and $y\otimes z$ degrees of freedom ($xy$) and $x\otimes y$ and $z$ degrees of freedom ($yz$), as well as the geometric non-separability, for the states given in Fig.~\ref{fig:OTHubb}.  In spite of the fact that properties, e.g.~the tunneling in the excited state, show significant characteristics of the non-separability of the potential, each eigenstate itself is very nearly separable, requiring a Schmidt rank of $\tilde{\chi}=5$ across any bipartition to capture the state in the 2-norm to an accuracy of $10^{-12}$ (see Eq.~\eqref{eq:SchmidtError}).  This near-separable character is also borne out in the separability measures of Fig.~\eqref{fig:OTHubb}.  For example, the geometric non-separability $\mathcal{E}$ demonstrates that the ground and excited states can be represented by their respective nearest separable approximations with fidelities of $\approx 0.9995$ and $\approx 0.9987$.  As a general rule, bound states of higher energy are less separable than bound states of lower energy.  Also, it is interesting to note that the degree of separability of the states of the double-well optical tweezer are not monotonic functions of the tweezer depth.

\begin{figure}[tp!]
\centerline{\includegraphics[width=0.8 \columnwidth]{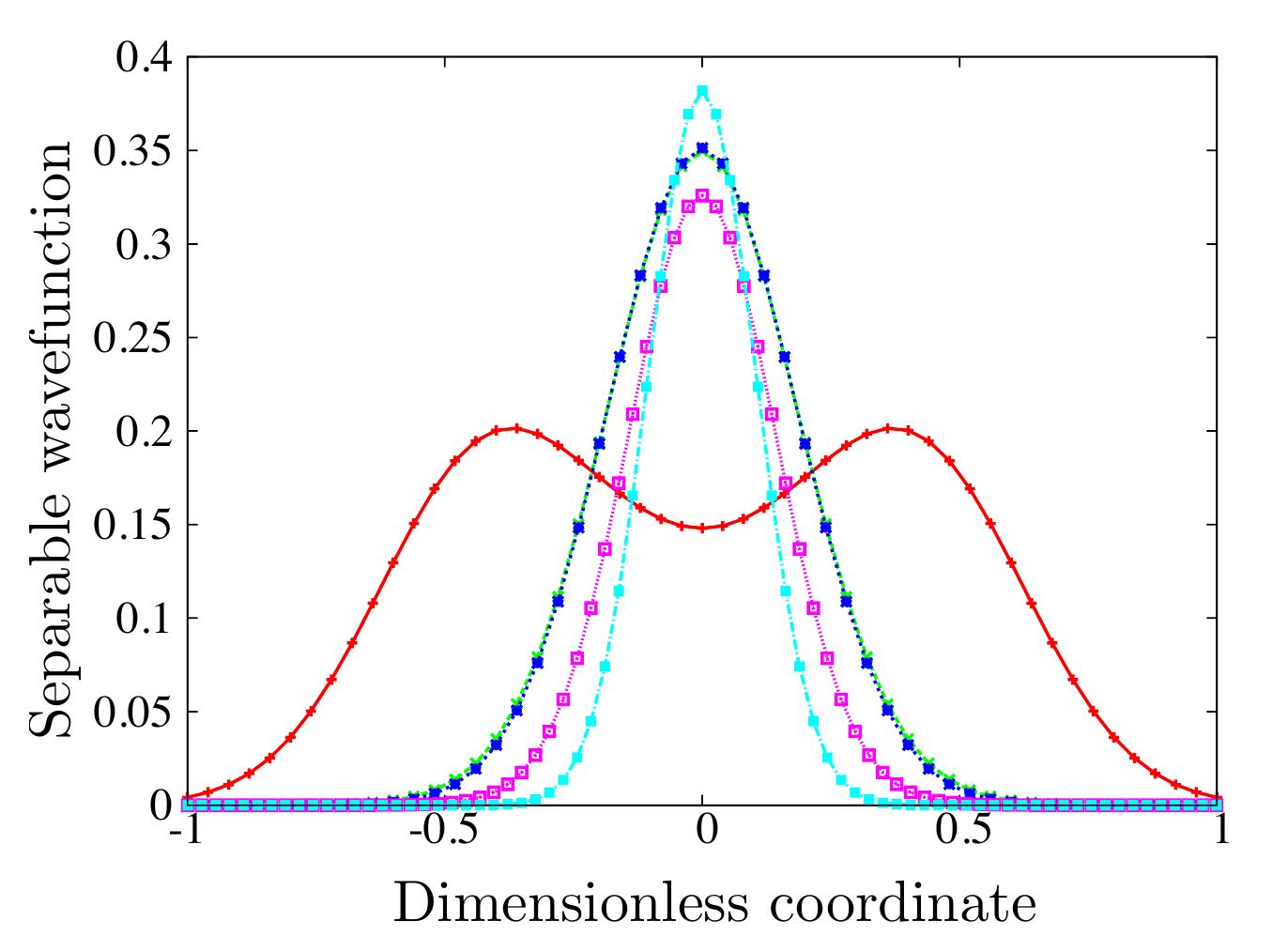}}
\caption{\label{fig:NSOT} 
\emph{Nearest separable states of a double-well optical tweezer and their harmonic approximations}.  The components of the nearest separable wavefunction $\tilde{X}(x/w_0)$ (red), $\tilde{Y}(y/w_0)$ (green), and $\tilde{Z}(z/z_R)$ (magenta) for a symmetric double-well optical tweezer with depth $V=52\;$kHz are shown as functions of their respective dimensionless coordinates.   The harmonic approximations are also shown for $y$ (dark blue) and $z$ (light blue), demonstrating significant anharmonicity in the $z$ degree of freedom.}
\end{figure}

The components of the nearest separable state $\psi\left(\mathbf{r}\right)=\tilde{X}(x)\tilde{Y}(y)\tilde{Z}(z)$, obtained using the method in Appendix~\ref{sec:MPSVari}, are shown in Fig.~\ref{fig:NSOT}.  In addition, the harmonic oscillator approximations for the $y$ and $z$ degrees of freedom, e.g., $Y_{\mathrm{ho}}(y)=\exp(-y^2/(2a_{\mathrm{ho}}^2))/\sqrt{a_{\mathrm{ho}}\sqrt{\pi}}$ with $a_{\mathrm{ho}}=w_0(2V/E_{w_0})^{-1/4}$, are shown for comparison.  The nearest separable state along the $y$ direction is accurately represented by its harmonic approximation, while the $z$ state is considerably wider than its harmonic counterpart and has a different shape, showing strong anharmonicity.  Finally, we stress that even though two states $\psi(\mathbf{r})$ and $\phi(\mathbf{r})$ may both be nearly separable, $\psi(\mathbf{r})\approx X_{\psi}(x)Y_{\psi}(y)Z_{\psi}(z)$ and $\phi(\mathbf{r})\approx X_{\phi}(x)Y_{\phi}(y)Z_{\phi}(z)$, this does not imply that $X_{\psi}(x)\approx X_{\phi}(x)$ etc..

\subsection{Optical lattice}

In this section, we turn our attention to a non-separable lattice of the form Eq.~\eqref{eq:VlattX}, where we will fix $V=12\,E_R$, with $E_R=\hbar^2 \pi^2/2m a^2$ the recoil energy of $^{87}$Sr in a lattice of spacing $a=406.72\;$nm, and $w_0=30\;\mu$m.  The ``running-wave" component of the lattice $V_{\mathrm{const}}$ will be left as a variable to show the effect of increasing transverse confinement on the axial motion.  The transverse confinement can be characterized by the effective oscillator frequency $\hbar\omega =\sqrt{8 E_{w_0} \left(V_{\mathrm{const}}+V\right)}$ resulting from a harmonic expansion near the Gaussian potential minimum.

\subsubsection{Construction of Wannier functions}

As discussed in Sec.~\ref{sec:QAB}, the states of the non-separable 1D optical lattice may be variationally obtained in the form
\begin{align}
\label{eq:fullOLpsi} \psi_{m q \nu}\left(\mathbf{r}\right)&=\frac{e^{im\phi}}{\sqrt{2\pi}} \sum_{n n_z}c_{\nu; m, n,q,n_z} \mathcal{R}_{m,n}\left(r\right) \varphi_{qn_z}\left(z\right)\, ,
\end{align}
where $\mathcal{R}(r)$ are a set of functions obtained from a DVR calculation using the transverse Gaussian potential and $\varphi_{qn_z}(z)$ are Bloch functions with quasimomentum $q$ and band index $n_z$ obtained from a plane-wave calculation.  Here, we briefly describe how we construct localized Wannier functions in this non-separable case by analogy with the separable case worked out by Kohn~\cite{Kohn_59}.  In Kohn's original scenario, which is an inversion-symmetric 1D lattice, real Wannier functions of maximal localization are obtained by using the transformation property of Bloch functions under inversion $\varphi_{q,n_z}(-z)=(-1)^{n_z+1}\varphi_{-q,n_z}(z)$ (we index $n_z$ starting from 1) and the requirement that the Bloch functions are smooth functions of $q$ to fix the gauge, i.e.~phase, ambiguity in the Bloch functions.  Using this choice of phases, maximally localized Wannier functions follow from
\begin{align}
w_{n_z i}(z)&=\frac{1}{\sqrt{L}}\sum_{q\in\mathrm{BZ}} e^{-i q z_i} \varphi_{q,n_z}(z)\, ,
\end{align}
where $L$ is the number of lattice sites and $z_i$ the centering site location.  The resulting Wannier functions transform under inversion as $w_{n_z,i}(-z)=(-1)^{n_z+1}w_{n_z,-i}(z)$: the Wannier function center is inverted and a phase may be acquired.  

In the non-separable case, we can generalize the inversion symmetry transformation to $\psi_{q \nu}\left({r},-z\right)=P_{\nu} \psi_{-q \nu}\left({r},z\right)$, where $P_{\nu}=1$ if the dominant weight of the state $\psi$ lies in bands $n_z=1,3,5,\dots$ and $-1$ otherwise.  This phase can also be set unambiguously at the BZ center $q=0$, where translations and inversions commute.  The remaining phase ambiguities are the Bloch function phases in $q\in(0,\pi/a)$, which become $\pm1$ under the requirement of real Wannier functions.  We fix the gauge here by requiring that the Bloch functions be smooth functions of $q$, by analogy with Kohn's original work.  With these phase conventions, Wannier functions are constructed from Eq.~\eqref{eq:fullOLpsi} as
\begin{align}
\label{eq:NSWannier}w_{m \nu i}\left(\mathbf{r}\right)&=\frac{1}{\sqrt{L}}\sum_{q\in\mathrm{BZ}}e^{-iqz_i}\psi_{m q \nu}\left(\mathbf{r}\right)\,.
\end{align}
Our construction produces Wannier functions which transform into each other up to phases with the symmetries of the lattice and reduce to the maximally localized Wannier functions of Kohn when the transverse motion separates from the axial motion.  We note that the particular choice of Wannier functions affects only interaction matrix elements and not tunneling matrix elements, as the latter are set by the energy as a function of $q$ alone, see Eq.~\eqref{eq:Jdisp}.

\begin{figure}[tp!]
\centerline{\includegraphics[width=0.8 \columnwidth]{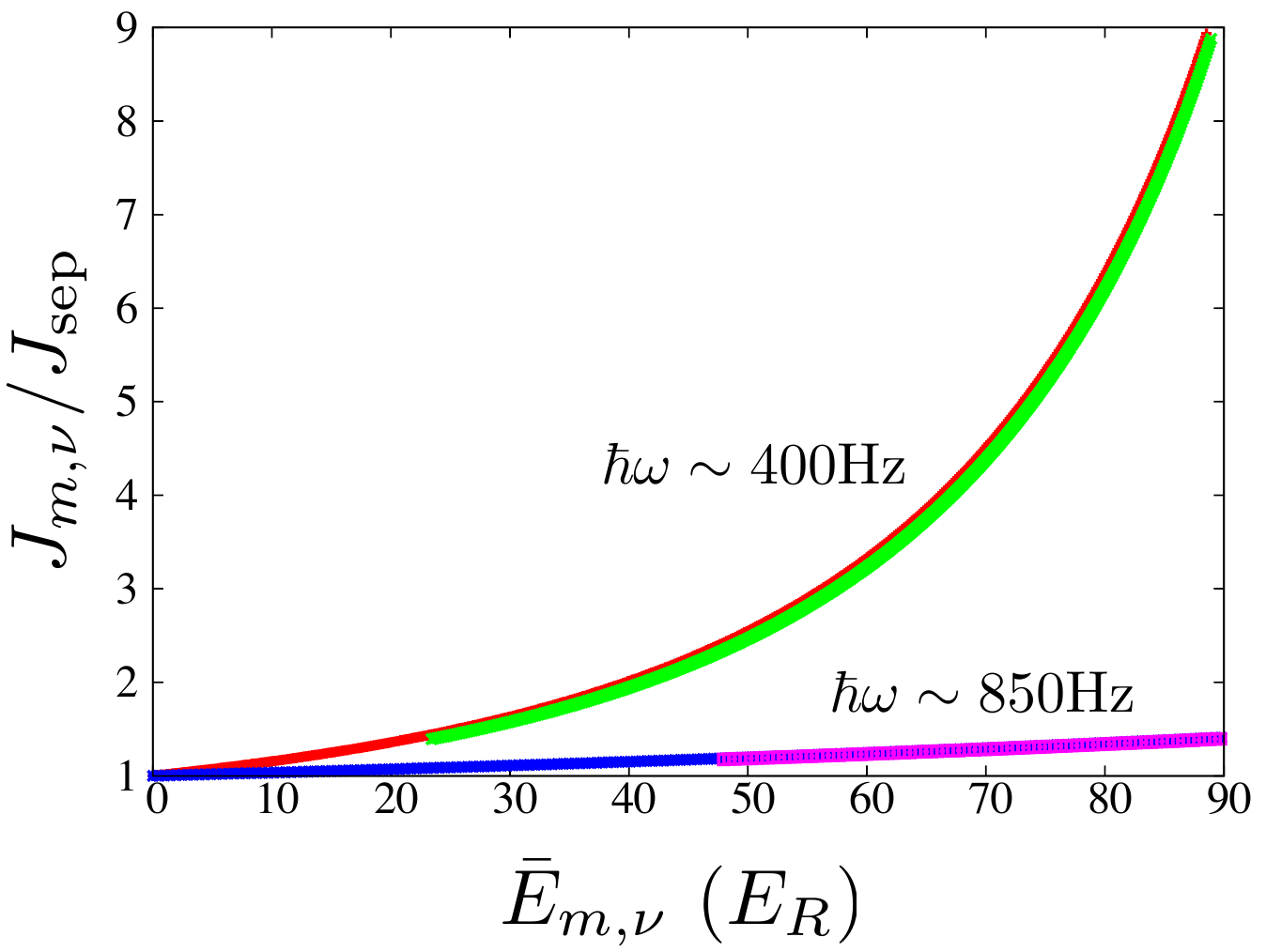}}
\caption{\label{fig:OLTunn} 
\emph{Tunneling vs.~Energy}. The tunneling, in units of the separable lattice tunneling, as a function of Wannier state energy for $V=12\,E_R$ and $w_0=30\;\mu$m.  The upper pair of curves corresponds to $V_{\mathrm{const}}=88\,E_R$, giving a transverse frequency $\hbar\omega\sim 400\;$Hz, with the red (green) curves corresponding to azimuthal quantum number $m=0$ ($m=200$).  The lower pair of curves has $V_{\mathrm{const}}=388\,E_R$ ($\hbar\omega\sim 850\;$Hz), with blue (magenta) corresponding to $m=0$ ($m=200$).  $J$ depends dominantly on the transverse mode energy, as intuitively understood through the lowest-order harmonic expansion of the transverse potential.}
\end{figure}

\subsubsection{Tunneling and interaction parameters}

Using the energy dispersion $E_{m,q,\nu}$ of state $\psi_{m q \nu}\left(\mathbf{r}\right)$, the tunneling of Wannier state $w_{m \nu i}\left(\mathbf{r}\right)$ between neighboring lattice sites is given by
\begin{align}
\label{eq:Jdisp}J_{m,\nu}&=-\frac{1}{L}\sum_{q\in\mathrm{BZ}}e^{-i q a}E_{m,q,\nu}\, ,
\end{align}
i.e.~the Fourier transform of the band structure in the relative lattice coordinate.  Similarly, the Wannier function energy, Eq.~\eqref{eq:onsiteE}, is given as the average of the dispersion in the BZ,
\begin{align}
\bar{E}_{m,\nu}&=\frac{1}{L}\sum_{q\in\mathrm{BZ}}E_{m, q,\nu}\, .
\end{align}
In Fig.~\ref{fig:OLTunn} we show the behavior of the lowest-band tunneling as a function of the Wannier state energy, with the former measured in units of the separable (1D) lattice lowest-band tunneling $J_{\mathrm{sep}}$ and the latter measured in units of the recoil energy, with the zero of energy being the ground state energy.  We define the lowest band as being the set of states such that $\sum_{n}c_{\nu;n,q,n_z}^2$ in Eq.~\eqref{eq:fullOLpsi} is maximal for $n_z=1$, where $n_z$ labels the ``bare" bands used in the variational expansion of Eq.~\eqref{eq:fullOLpsi}.  For the parameters we consider, the mixing between the bare bands is slight ($\ge 85\%$ of the population is in $n_z=1$ for the energy range we consider) and there is no ambiguity in this definition.  Hence, the energy on the $x$-axis of Fig.~\ref{fig:OLTunn} correlates dominantly to transverse mode energy.  The tunneling generally increases with increasing energy, which can be understood by expanding Eq.~\eqref{eq:VlattX} to lowest order in $r$
\begin{align}
\nonumber V_{\mathrm{latt}}\left(r,z\right)&=V_{\mathrm{const}}\left(-1+2\frac{r^2}{w_0^2}\right)\\
&+V\left(-1+2\frac{r^2}{w_0^2}\right)\cos^2\left(k z\right)\, .
\end{align}
Hence, to lowest order, the lattice potential depth is lowered by an amount proportional to $\langle r^2\rangle$, which is itself proportional the transverse mode energy in the harmonic oscillator approximation.  Semiclassically, a particle in a higher transverse mode spends more time near the classical turning points, and here the potential depth is smaller due to the non-separability.  Because of the non-linear relationship between lattice depth and tunneling, the relationship between transverse mode energy and tunneling is also non-linear.  The semiclassical reasoning for the dependence of tunneling on energy is also supported by comparing the red and green curves in Fig.~\ref{fig:OLTunn}, which correspond to the $m=0$ and $m=200$ states, respectively.  The tunneling is well-represented by a function only of the transverse mode energy, even when the nature of the excitation (azimuthal or radial) is very different.

\begin{figure}[tp!]
\centerline{\includegraphics[width=1.0 \columnwidth]{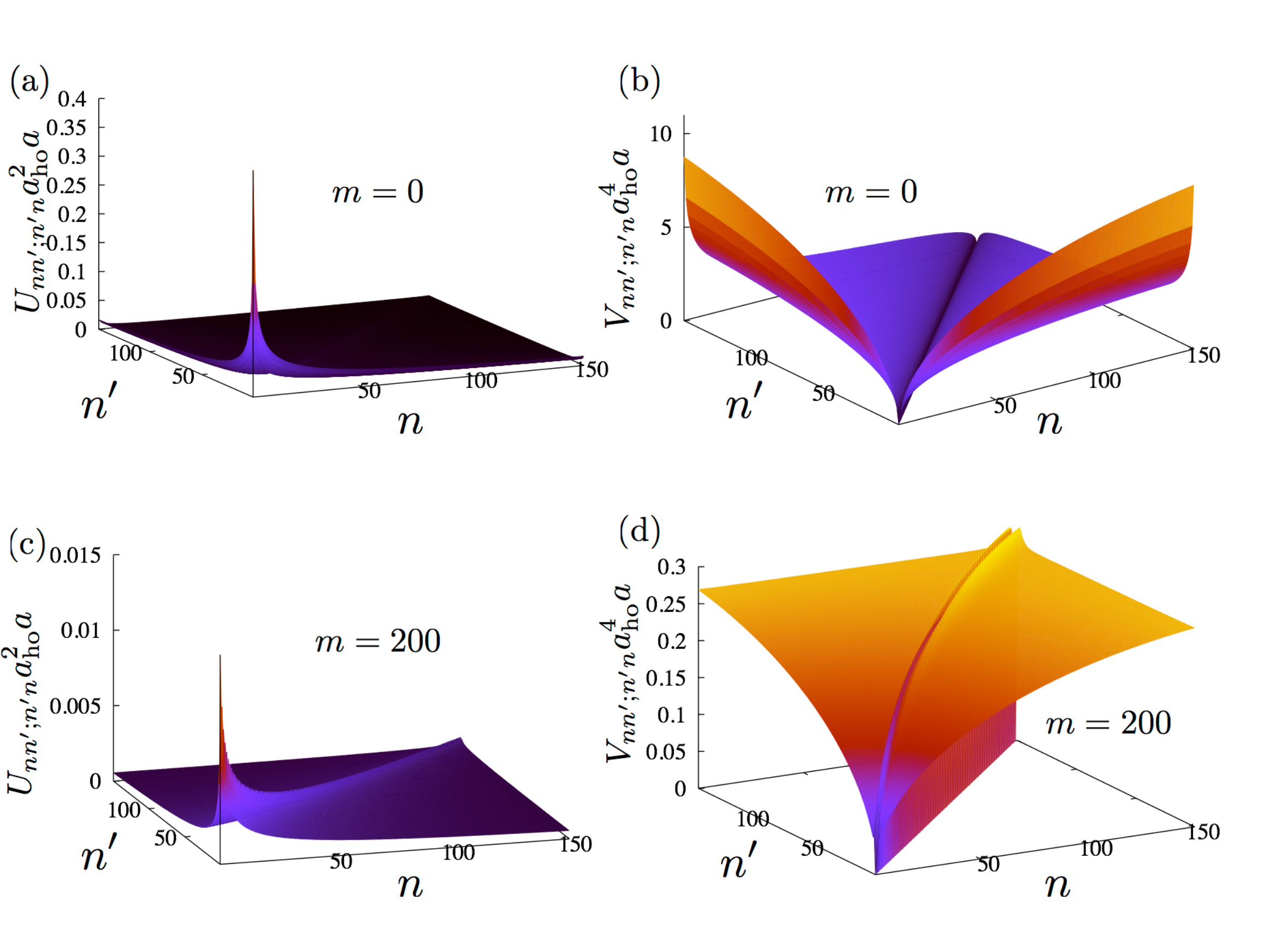}}
\caption{\label{fig:InteractionPlot} 
\emph{Interaction matrix elements in a non-separable optical lattice} The $s$- and $p$-wave interaction integrals Eqs.~\eqref{eq:swavepp}-\eqref{eq:pwavepp}, for $m=0$ (top row) and $m=200$ (bottom row).  $n$ and $n'$ are eigenstate indices, which correlate to transverse mode quantum numbers.  $s$-wave ($p$-wave) interactions show a slow decay (growth) with transverse mode energy.}
\end{figure}

The red and green curves in Fig.~\ref{fig:OLTunn} correspond to $V_{\mathrm{const}}=88\,E_R$, giving a transverse confinement harmonic oscillator frequency of $\sim 400\;$Hz.  In this case, the tunneling changes by nearly an order of magnitude in the given energy range, which can amount to a significant thermal dependence of the effective tunneling amplitude.  As a point of comparison, for strontium in a magic lattice, $10\,E_R$ corresponds to a temperature of $\approx 1.66\mu $K, which is comparable to the operating temperatures of optical lattice clocks~\cite{Martin09082013,Hinkley13092013}.  The blue and magenta curves, which correspond to the $m=0$ and $m=200$ states, are computed for $V_{\mathrm{const}}=388\,E_R$, giving a transverse confinement harmonic oscillator frequency of $\sim 850\;$Hz.  The dependence of the tunneling on transverse mode energy in the same energy range is now significantly smaller, less than a factor of two.  Hence, the axial motion, represented through the tunneling properties of the lattice, can be effectively decoupled from the transverse motion by increasing the effective transverse confinement frequency.  

In Fig.~\ref{fig:InteractionPlot} we show the $s$- and $p$-wave integrals Eqs.~\eqref{eq:swavepp}-\eqref{eq:pwavepp}.  We use units $a_{\mathrm{ho}}^{(\ell+1)2}a$, with $\ell=0$ (1) for $s$- ($p$-) wave interactions, where $a_{\mathrm{ho}}$ is the harmonic length associated with the radial direction, in order to facilitate comparison with previous works~\cite{Martin09082013,Rey2014311} using a harmonic oscillator approximation.  Similar to the case using the harmonic approximation, we find a slow decay of $s$-wave interactions with mode energy, and a slow growth of $p$-wave interactions with increasing mode energy.  When thermally averaged, this weak energy dependence leads to quantitatively similar behavior between the Gaussian, non-separable trap case and its harmonic oscillator approximation when the temperature is comparable or larger than the transverse mode spacing.


\subsection{Non-separability and anharmonicity of optical lattice states}

\begin{figure}[tp!]
\centerline{\includegraphics[width=0.8 \columnwidth]{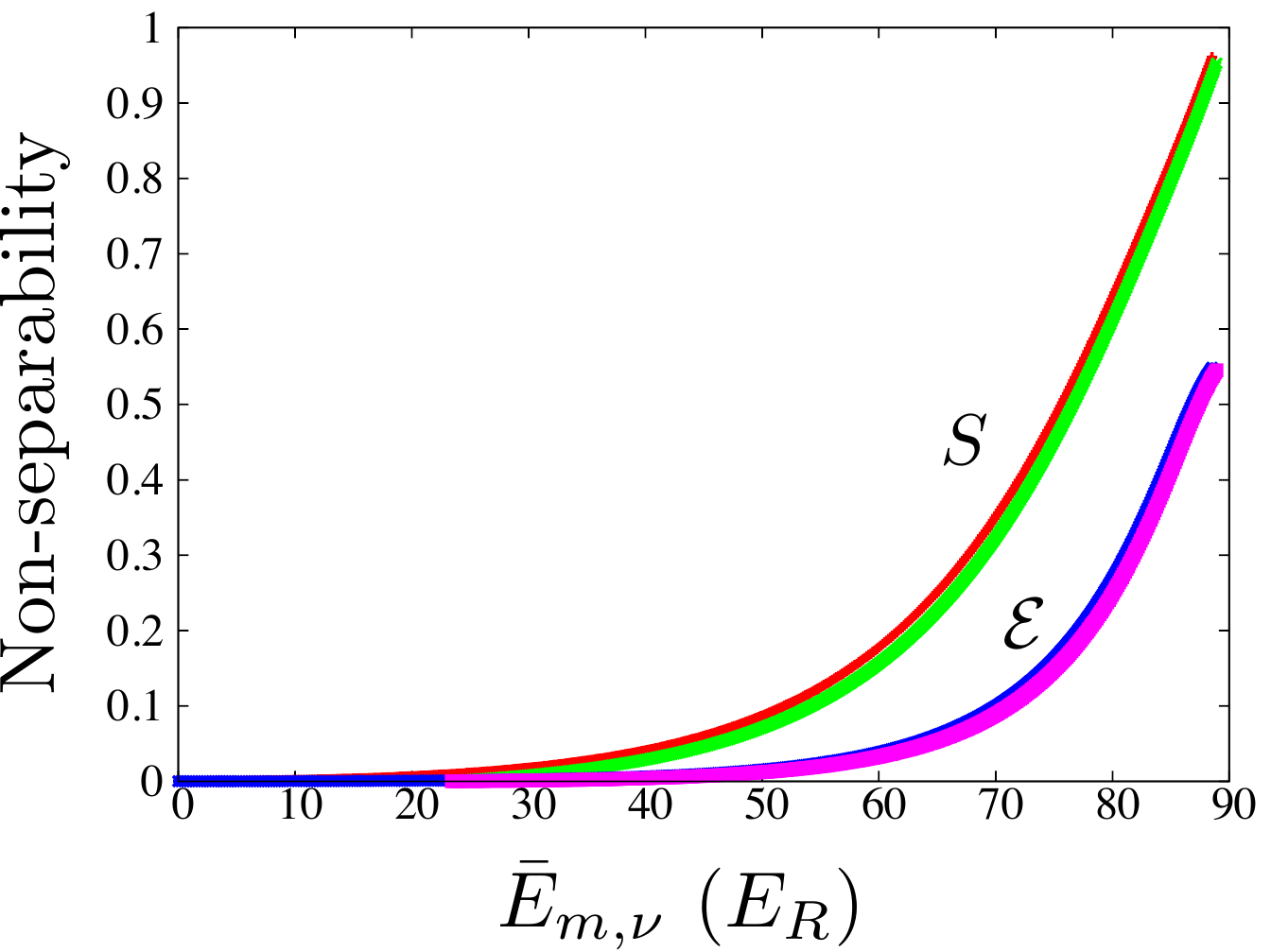}}
\caption{\label{fig:OLVNE} 
\emph{Non-separability of 1D optical lattice eigenstates} The von Neumann entropy of non-separability $S$ and the geometric non-separability $\mathcal{E}$ as a function of Wannier state energy for $V_{\mathrm{const}}=88\,E_R$.  Red and blue (green and magenta) curves correspond to states of azimuthal quantum number $m=0$ ($m=200$).  As was the case for the tunneling (Fig.~\ref{fig:OLTunn}), the non-separability is dominantly a function of transverse mode energy.}
\end{figure}

We now consider the quantitative non-separability of the eigenstates of Eq.~\eqref{eq:VlattX}.  We decompose a band of states as
\begin{align}
\label{eq:psiBNS}\psi_{m q \nu}\left(\mathbf{r}\right)&=\sum_{\mu}R^{\left(m\right)}_{\mu}(r)Z^{\left(m\right)}_{\mu, q}\left(z\right)\, .
\end{align} 
The non-separability measures for this decomposition are shown as a function of Wannier state energy for $V_{\mathrm{const}}=88\,E_R$ in Fig.~\ref{fig:OLVNE}.  While the states near the bottom of the trap are nearly separable, more highly excited states can be very significantly non-separable.  As was also found for the tunneling, the non-separability is dominantly a function of the transverse mode energy, irrespective of its character (azimuthal or radial), as can be seen by comparing the red and blue curves (corresponding to $m=0$) with the green and magenta curves ($m=200$).  

\begin{figure}[tbp]
\centerline{\includegraphics[width=0.7 \columnwidth]{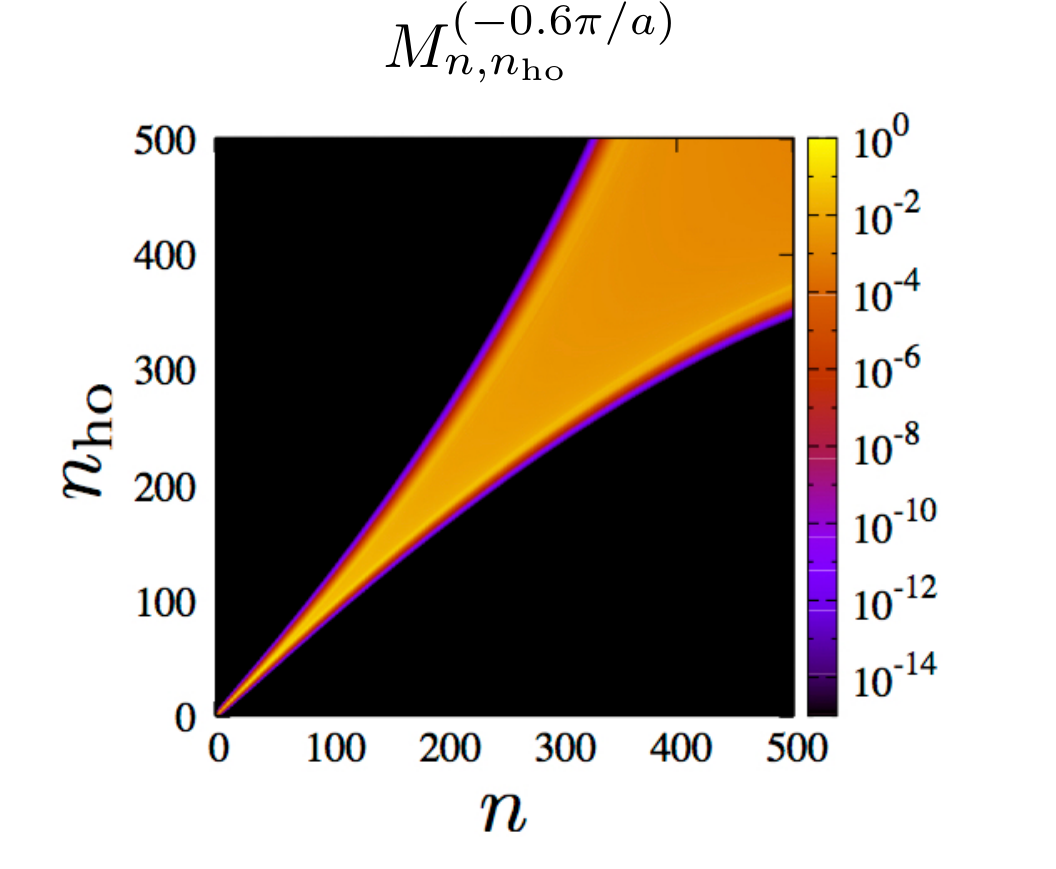}}
\caption{\label{fig:LAH} 
\emph{Anharmonicity of non-separable optical lattice states}.  The overlap of the radial part of the wavefunction, indexed by $n$, with the approximate harmonic oscillator wavefunctions, indexed by $n_{\mathrm{ho}}$, shows that the radial wavefunction is spread across many harmonic oscillator modes in excited states.  The values at quasimomentum $q=-0.6\pi/a$ are representative of general quasimomentum away from high-symmetry points.}
\end{figure}

To quantitatively assess the degree of anharmonicity of the optical lattice eigenstates, we compute the overlap of the radial part of these states with harmonic oscillator functions chosen to match the local curvature of the potential.  Namely, we compute 
\begin{align}
M_{\nu,n_{\mathrm{ho}}}^{\left(q\right)}&=\sum_{n_z,n} c_{\nu; n,q,n_z}^2 \left[\int dr \mathcal{R}_{n}(r)\phi_{n_{\mathrm{ho}}}(r)\right]^2
\end{align}
in the notation of Eq.~\eqref{eq:fullOLpsi}.  In the lowest-order harmonic expansion of the potential, $M_{\nu,n_{\mathrm{ho}}}^{\left(q\right)}=\delta_{\nu,n_{\mathrm{ho}}}$.  The function $M_{n,n_{\mathrm{ho}}}^{\left(-0.6\pi/a\right)}$ is shown in Fig.~\ref{fig:LAH} for $V_{\mathrm{const}}=88\,E_R$, where $n$ is the approximate radial quantum number.  The values at quasimomentum $q=-0.6\pi/a$ are representative of general quasimomentum away from high-symmetry points.  For low-lying eigenstates,  $M_{n,n_{\mathrm{ho}}}^{\left(-0.6\pi/a\right)}\approx \delta_{n,n_{\mathrm{ho}}}$, showing the near-harmonic nature of these states.  However, in higher-lying eigenstates, the character of the wavefunction is spread over many harmonic oscillator modes, demonstrating significant anharmonicity.



\section{Conclusions and outlook}
\label{sec:Concl}

We have studied the properties of particles confined in non-separable 3D optical potentials, focusing in particular on a double-well optical tweezer array and a 1D optical lattice with transverse Gaussian confinement.  Our main methodology was discrete variable representations (DVRs), which couple the rapid convergence of spectral methods with the flexibility and simplicity of grid-based methods, as well as variational methods built on top of DVR approaches.  We found that parameters relevant to the construction of effective many-body models, such as tunneling and interaction matrix elements, can be significantly different from their separable counterparts.  In particular, we found that lowest-band tunneling amplitudes in a non-separable lattice can be increased by nearly an order of magnitude for a state with significant transverse mode energy compared to a state in the transverse ground state.  Similarly, we found the lowest-lying state with an excitation transverse to the tunneling direction in a double-well optical tweezer has a tunneling rate $\approx 20\%$ larger than the ground state.  The fact that tunneling depends on transverse motional state could have a range of applications, such as thermometry and quantum simulation of multi-component systems with effective mass imbalance.  Interactions were found to be less sensitive to non-separability and anharmonicity compared with tunneling amplitudes.

In addition to discussing how effective model parameters change with the trap variables, we also presented a quantitative analysis of the non-separability of individual eigenfunctions by adapting methods from the theory of matrix product states (MPSs).  In particular, we developed a canonical form for non-separable states in terms of a contraction of low-rank tensors describing motion along each independent direction, and discussed how this canonical MPS form is useful for storage, computation, visualization, and quantification of non-separability.  Based on this canonical form, we discussed three measures of non-separability: the Schmidt rank and von Neumann entropy of non-separability, both of which depend on a specific bipartition of degrees of freedom, and the geometric non-separability, a global measure of non-separability that quantifies the distance to the nearest separable state.  Finally, we also presented a variational algorithm for determining the nearest separable state to a given state, and showed how this nearest separable state can be used to gain intuition about anharmonicity, to classify quantum states, and to assess the convergence of algorithms.  We found that the low-lying states of the non-separable potentials we considered were nearly separable despite often being significantly anharmonic.  This observation strongly motivates the use of the MPS canonical form for non-separable states as a ansatz that can be variationally optimized at fixed degree of non-separability with significantly reduced computational resources compared to direct diagonalization.

We would like to acknowledge useful discussions with Michael Foss-Feig, Adam Kaufman, Brian Lester, Cindy Regal, and Xibo Zhang.  This work was supported by JILA-NSF-PFC-1125844, NSF-PIF-1211914, ARO, AFOSR, and AFOSR-MURI.  MLW acknowledges support from the NRC postdoctoral fellowship program.

\appendix 
\section{Interaction parameters for radial functions}
\label{sec:RInteractions}

As mentioned in Sec.~\ref{sec:BDVR}, the Bessel DVR grids are different for different angular momentum $m$, and so there is no single DVR quadrature which can be used to integrate products of functions which have different values of $m$.  Hence, in order to find interaction matrix elements between radial functions, we must use some other basis set.  A natural choice is to use the sinc DVR in $r$, adapted for odd-parity potentials to ensure that all radial functions vanish at $r=0$.  Using the sinc DVR, all radial functions are expressed in terms of the same grid, and so the sinc DVR quadrature can be used to find interaction matrix elements.  This procedure works well for $\left|m\right|\ge 3$, but is slowly convergent for $\left|m\right|\le 2$.  This is due to the fact that the sinc DVR basis functions do not have the appropriate asymptotic behavior near $r=0$.  As $|m|$ increases, the wave function is very small near this region due to the centrifugal potential, and so the mismatch of boundary conditions does not affect the convergence of the algorithm.  For the states with $|m|\le 2$, we use the following alternate procedure:
\begin{enumerate}
\item Solve for the eigenstates of the Gaussian potential using the Bessel DVR.
\item Solve for the eigenstates of a radial harmonic oscillator with the same local curvature near the potential minimum as the Gaussian potential using the Bessel DVR.
\item Find the expansion of the Gaussian potential states in terms of the harmonic oscillator states using the Bessel DVR quadrature.
\item Solve for the eigenstates of a Cartesian harmonic oscillator with the same frequency as the radial oscillator using the sinc DVR.
\item Use the Cartesian harmonic oscillator functions to obtain the radial harmonic oscillator functions on an equally spaced radial grid using transformation Eq.~\eqref{eq:RadialCartTrans} below.
\item Using the results of step 3 and 5, find the values of the Gaussian potential states on an equally spaced radial grid.
\end{enumerate}
One may object that steps 3-5 are unnecessary, as the exact wave functions for the harmonic oscillator are known.  However, evaluation of these wavefunctions directly in terms of orthogonal polynomials is numerically unstable.  In contrast, using DVR-based eigenvalue methods is numerically stable, even for very highly excited states.

The expansion of the radial harmonic oscillator states $|m,n_r\rangle$ in terms of Cartesian harmonic oscillator states $|p\rangle|q\rangle$, where ``$p$" labels the $x$ harmonic oscillator state and ``$q$" labels the $y$ harmonic oscillator state may be accomplished as
\begin{align}
\label{eq:RadialCartTrans} &|m,n_r\rangle=\sum_{k=-n_r-|m|/2}^{n_r+|m|/2}\left(-i\right)^{(n_r+|m|/2)-k}\left(-1\right)^{n_r}\\
\nonumber &\times d^{\left(n_r+|m|/2\right)}_{m/2,k}\left(-\frac{\pi}{2}\right)|n_r+\frac{|m|}{2}+k\rangle|n_r+\frac{|m|}{2}-k\rangle\, ,
\end{align}
where $d^{\left(\ell\right)}_{m m'}\left(\theta\right)$ is the Wigner little-$d$ matrix~\cite{Interbasis}.  Since we do not need the full 2D wavefunction but only the radial function, we can consider $|m,n_r\rangle$ along the line of polar angle $\phi=0$ in which $r=x$.  Here, we have
\begin{align}
&\langle r,\phi=0|m,n_r\rangle=\sum_{k=-n_r-|m|/2}^{n_r+|m|/2}\left(-i\right)^{(n_r+|m|/2)-k}\left(-1\right)^{n_r}\\
\nonumber &\times d^{\left(n_r+|m|/2\right)}_{m/2,k}\left(-\frac{\pi}{2}\right)\langle x|n_r+\frac{|m|}{2}+k\rangle\langle 0|n_r+\frac{|m|}{2}-k\rangle\, .
\end{align}
Due to the fact that $\langle x|n\rangle$ is an odd (even) function if $n$ is odd (even), only $k$ such that $(n_r+\frac{|m|}{2}-k)$ is even contribute to the sum, and hence the radial wavefunction is real.  In contrast to the recurrence relations required for the harmonic oscillator wavefunctions themselves, $d$ matrices have a numerically stable recurrence relation~\cite{DachselDmats}.

At the end of this procedure, we have the values of the radial functions evaluated on a grid of equally spaced radial points, and so we could think of this as being an expansion of the radial wave functions in terms of an odd-parity sinc DVR basis and use the associated quadrature to evaluate overlaps and derivatives.  However, due to the mismatch in boundary conditions between the radial function and the sinc DVR functions, such an expansion is inaccurate, and instead standard grid-based techniques for integration, e.g.~Simpson's rule, and evaluation of derivatives by high-order finite differencing will yield more accurate results.  In contrast to performing derivatives and integration with DVR quadrature, where the calculation converges exponentially fast, this procedure features only algebraic convergence with the step size $\Delta x$, with the particular rate of convergence set by the finite differencing and integration scheme. 

\section{Obtaining the MPS form of a non-separable state and a variational algorithm for finding the nearest separable state}
\label{sec:MPSVari}

Similar to the case of the Schmidt form, the MPS representation of a quantum state can be obtained via the singular value decomposition (SVD).  In particular, let us assume we have a state $\psi_{x_i,y_j,z_k}$, where $x_i$, $y_j$, and $z_k$ are discrete indices running over some finite set of basis functions (e.g. DVR functions).  Then, we can obtain the representation Eq.~\eqref{eq:CartMPS} as
\begin{align}
\label{eq:MPSStart}\psi_{(x_i,y_j),z_k}&\;\overrightarrow{\mathrm{SVD}}\; \sum_{\nu}U_{\left(x_i,y_j\right),\nu}S_{\nu} V_{\nu,z_k}\\
\label{eq:MPSyzdef}Z_{\mu_{yz}}\left(z_k\right)&=V_{\mu_{{yz}},z_k}\, ,\;\; A_{x_i,\left(y_j,\nu\right)}=U_{\left(x_i,y_j\right),\nu}S_{\nu}\\
A_{x_i,\left(y_j,\nu\right)}&\;\overrightarrow{\mathrm{SVD}}\; \sum_{\mu} U_{x_i,\mu}S_{\mu}V_{\mu,\left(y_j,\nu\right)}\\
\label{eq:MPSDone}X_{\mu_{xy}}\left(x_i\right)&=U_{x_i,\mu_{{xy}}}\, ,\;\; Y_{\mu_{xy}\mu_{yz}}(y_j)=S_{\mu_{xy}}V_{\mu_{xy},\left(y_j,\mu_{yz}\right)}\, .
\end{align}
Here, $\left(a,b\right)$ denotes the Kronecker product of the indices $a$ and $b$, and $\overrightarrow{\mathrm{SVD}}$ denotes matrix decomposition of the left hand side into the right hand side via the SVD.  The particular form of Eq.~\eqref{eq:CartMPS} obtained with Eqs.~\eqref{eq:MPSStart}-\eqref{eq:MPSDone} is a mixed canonical form~\cite{Schollwoeck_11} with the gauge conditions $\int dx X_{\mu}\left(x\right) X_{\nu}\left(x\right)=\delta_{\mu,\nu}$, $\int dz Z_{\mu}\left(z\right) Z_{\nu}\left(z\right)=\delta_{\mu,\nu}$, and $\sum_{\mu \nu} \int dy Y_{\mu\nu}(y)^2=1$.

Finding the nearest separable state for a multi-partite system is not as simple as for the two-partite case~\cite{1367-2630-12-2-025008, Wei_Thesis}.  However, given the MPS form of a non-separable state Eq.~\eqref{eq:CartMPS}, a variational algorithm for obtaining the nearest separable state $\langle \mathbf{r}|\psi_{\mathrm{sep}}\rangle=\tilde{X}\left(x\right)\tilde{Y}\left(y\right)\tilde{Z}\left(z\right)$ can be devised by optimizing the tensors $\tilde{X}$, $\tilde{Y}$, and $\tilde{Z}$ individually in a round-robin fashion.  Such an alternating least-squares algorithm is similar to the local energy optimization coupled with sweeping over lattice sites used in the DMRG algorithm of condensed matter physics~\cite{Schollwoeck_11}.  The optimal local tensor updates in the case at hand are
\begin{align}
\label{eq:Xupdate}\tilde{X}\left(x\right)&=\sum_{\mu_{xy}\mu_{yz}}X_{\mu_{xy}}\left(x\right) \left(Y_{\mu_{xy}\mu_{yz}}\cdot \tilde{Y}\right)\left(Z_{\mu_{yz}}\cdot \tilde{Z}\right)\, ,\\
\tilde{Y}\left(y\right)&=\sum_{\mu_{xy}\mu_{yz}}\left(X_{\mu_{xy}}\cdot \tilde{X}\right) Y_{\mu_{xy}\mu_{yz}}\left(y\right)\left(Z_{\mu_{yz}}\cdot \tilde{Z}\right)\, ,\\
\label{eq:Zupdate}\tilde{Z}\left(z\right)&=\sum_{\mu_{xy}\mu_{yz}}\left(X_{\mu_{xy}}\cdot \tilde{X}\right)\left(Y_{\mu_{xy}\mu_{yz}}\cdot \tilde{Y}\right) Z_{\mu_{yz}}\left(z\right)\, ,
\end{align}
where the updated tensor, e.g., $\tilde{X}\left(x\right)$ is normalized after the update and $\tilde{A}\cdot A$ is shorthand for $\int d\xi \tilde{A}(\xi)A(\xi)$.  Convergence can be assessed by stationarity of the functional $\left||\psi_{\mathrm{sep}}\rangle-|\psi\rangle\right|^2$.  An appropriate starting guess is $\tilde{X}\left(x\right)=X_1\left(x\right)$, $\tilde{Y}\left(y\right)=Y_{1,1}\left(y\right)/\sqrt{Y_{1,1}\cdot Y_{1,1}}$, $\tilde{Z}\left(z\right)=Z_1\left(z\right)$, which would be the expectation for the nearest separable state based on applying the optimal truncation at each bipartition.  This algorithm can also be straightforwardly generalized to find the nearest state with fixed Schmidt rank non-separability $\tilde{\chi}_{xy}$ and $\tilde{\chi}_{yz}$.

{\bibliographystyle{prsty}}
{\bibliography{DVRRefs}}

\end{document}